\documentclass[twocolumn,amsmath,amssymb,pra]{revtex4-2}
\usepackage{graphicx}
\usepackage{amsmath,amsfonts,amssymb,amsthm}
\usepackage{fancyhdr}
\usepackage{mathrsfs}
\usepackage{epstopdf}
\usepackage{setspace}
\usepackage{bm}
\usepackage[colorlinks]{hyperref}
\usepackage{xcolor}
\usepackage{dcolumn}

\usepackage{tikz}
\usetikzlibrary{quantikz}
\graphicspath{{figures/}}

\begin{document}
\title{A quantum algorithm for solving open system dynamics on quantum\\ computers using noise}


\author{Juha Lepp\"akangas}
\affiliation{HQS Quantum Simulations GmbH, Rintheimer Str. 23, 76131 Karlsruhe, Germany}

\author{Nicolas Vogt}
\affiliation{HQS Quantum Simulations GmbH, Rintheimer Str. 23, 76131 Karlsruhe, Germany}

\author{Keith R. Fratus}
\affiliation{HQS Quantum Simulations GmbH, Rintheimer Str. 23, 76131 Karlsruhe, Germany}

\author{Kirsten Bark}
\affiliation{HQS Quantum Simulations GmbH, Rintheimer Str. 23, 76131 Karlsruhe, Germany}

\author{Jesse A. Vaitkus}
\affiliation{HQS Quantum Simulations GmbH, Rintheimer Str. 23, 76131 Karlsruhe, Germany}

\author{Pascal Stadler}
\affiliation{HQS Quantum Simulations GmbH, Rintheimer Str. 23, 76131 Karlsruhe, Germany}

\author{Jan-Michael Reiner}
\affiliation{HQS Quantum Simulations GmbH, Rintheimer Str. 23, 76131 Karlsruhe, Germany}

\author{Sebastian Zanker}
\affiliation{HQS Quantum Simulations GmbH, Rintheimer Str. 23, 76131 Karlsruhe, Germany}

\author{Michael Marthaler}
\affiliation{HQS Quantum Simulations GmbH, Rintheimer Str. 23, 76131 Karlsruhe, Germany}

\begin{abstract}
In this paper we present a quantum algorithm that uses noise as a resource.
The goal of our quantum algorithm is the calculation of operator averages of an open quantum system evolving in time.
Selected low-noise {\it system qubits} and noisy {\it bath qubits} represent the system and the bath of the open quantum system.
All incoherent qubit noise can be mapped to bath spectral functions.
The form of the spectral functions can be tuned digitally,
allowing for the time evolution of a wide range of open-system models at finite temperature.
We study the feasibility of this approach with a focus on the solution of the spin-boson model
and assume intrinsic qubit noise that is dominated by damping and dephasing.
We find that classes of open quantum systems exist where our algorithm performs very well, even with gate errors as high as 1\%.
In general the presented algorithm performs best if the system-bath interactions can be decomposed into native gates.
\end{abstract}

\maketitle


\section{Introduction}

Quantum computers promise a substantial speedup for solving certain types of numerical tasks,
in particular the simulation of large quantum systems~\cite{Georgescu2014}.
However, due to the large error rates and short coherence times of present quantum computers~\cite{Barthi2022},
only small examples have been demonstrated using digital quantum computing.
For useful near-term applications,
there is a need for current research to be focused on more efficiently exploiting noisy intermediate-scale quantum (NISQ) computers.
In this paper we present a quantum algorithm which, in fact, utilizes typical noise in NISQ devices by incorporating it into the computation itself.

Models for open quantum systems~\cite{Breuer2002, Weimer2021} (short, open-system models or OSM)
have been developed for the study of a small number of degrees of freedom, the system,
interacting with a large environment, the bath.
Accordingly, the use of an open-system model is often called the system-bath approach.
The idea is that certain dynamical or steady-state properties of the system can be the key to understanding the overall behavior of the full system-environment pair.
The bath can be modeled with less accuracy, making the approach economic.
One widely studied example is the energy transfer involved in photosynthesis~\cite{Ishizaki2012},
where the system is a finite number of local excitations (excitons) and the bath are the vibrational modes.
The effect of the bath on the system is described by a spectral function and
the effect strongly depends on its height and form~\cite{Spin_Boson_Rev}.
The limit of a smooth spectral function (on the scale of the system-bath couplings)
can be described by the well-known Lindblad or Bloch-Redfield master equation.

An open system is characterized by non-unitary time evolution.
Therefore, a quantum algorithm that time evolves an open-system model on a quantum computer must implement this in some way.
This can be accomplished by introducing entanglement with additional (external) qubits and performing measurements on them. 
Here, one class of methods implements directly a quantum map between an initial and final density matrix of the system~\cite{Perez2020,Head2021,Rost2020, Sun2021, Suri2022}.
This assumes that the time evolution is known beforehand. If one must instead solve for this time evolution,
a Trotterization needs to be implemented~\cite{Lloyd2001,Wang2011,Mi2023,Han2021,Hu2022,Barreiro2011,Rost2021,Kamakari2022,Guimaraes2023}.
Most such approaches consider an open-system model that is described by a Lindblad master equation.
A Trotter time evolution under a Lindblad master equation can be performed on a quantum computer using the same approach known for closed quantum systems:
the system is coupled to the external qubits at each Trotter step, which are measured and thus create non-unitary quantum operations.
Without access to quantum feedback control,
a technical challenge here is the required reset of the external qubits after each Trotter step during the computation.
This can however also be realized via controlled qubit dissipation~\cite{Barreiro2011, Han2021, Rost2021, Mi2023}.
Classical gates can also be implemented using similar approaches~\cite{Zapusek2022}.


One attractive way to implement open system time-evolution on NISQ computers is 
to directly map intrinsic qubit noise to noise processes in the model to be simulated~\cite{Lloyd1996, Tseng2010, Rost2020, Sun2021, Guimaraes2023}.
Here, 
noise is no longer an impediment, but is rather a key component of the computation itself.
Such an approach is more intuitive for analog quantum simulation. However, it can also be developed for digital quantum simulation;
although a Trotter time-evolution might involve many different gate operations,
it is possible to describe the effect of noise in the form of a static Lindblad master equation~\cite{Fratus2022}. 
One limitation of previous noise-utilizing algorithms has been that they fall in the category of smooth spectral functions.

In this paper, we present a noise-utilizing quantum algorithm that time evolves an open-system model that has a strongly structured bath.
The structured bath is represented by a finite number of bath modes subjected to Lindblad dissipation, a method known from classical numerical methods~\cite{Arrigoni2013, Dorda2014, Tamascelli2018, Chen2019, Pleasance2020, Pleasance2021}.
On a quantum computer, low and high noise qubits 
represent the system and the bath modes of the open quantum system.
Significant variations in qubit coherence times may appear in state-of-the-art quantum computers~\cite{GoogleSupremacySI},
and it is typically possible during calibration to make some qubits better at the expense of others.
In our work, we use this to our advantage by using the lower quality qubits as bath qubits
for mimicking the effect of a continuous spectral function.
By utilizing the intrinsic noise, we also avoid the need to introduce additional external qubits and their reset methods.
We give a detailed description of how the spectral function, as seen by the time-evolved system, can be made strongly structured, digitally tunable
and to match to the open-system model of interest.

The performance of the proposed algorithm is studied through numerical simulation on conventional computers.
In particular,
we study the quality of the results of the algorithm for a spin coupled strongly to a broad bosonic mode as well as to a bosonic ohmic bath.
Here, we establish a mapping between the open-system model and the noisy-algorithm model in the case of noiseless system-qubits.
We also study the solution for an electron-transport model by representing it using a generalized spin-boson model.
Here, we map also the system-qubit noise to the model spectral function.
Our central finding is that
the quantum algorithm performs best if the system-bath interactions can be decomposed into native gates,
for instance the XX Ising interaction to variable M{\o}lmer-S{\o}rensen (MS) gate or the XX + YY interaction to variable iSWAP gate.
The restriction to native gates can, however, be lifted for some open-system models.

The paper is structured as follows.
In Sec.~\ref{sec:SimulatedModel}, we introduce the concrete open system model
whose noise-utilizing quantum algorithm will be presented in this paper.
In Sec.~\ref{sec:BathMapping}, we present the protocol of the bath mapping, {\it i.e.},
the mapping between an open-system model and a noisy-algorithm model.
In Sec.~\ref{sec:examples}, we go through 
three practical examples of solving the open-system dynamics.
The examples were implemented using numerical simulations on conventional computers.
Conclusions and discussion are given in Sec.~\ref{sec:discussion}.
Many important details of our approach are left to be presented in the Appendices.
In Appendix~\ref{Appendix:multi_spin_systems}, we generalize the approach to also cover multi-spin systems.
In Appendix~\ref{Appendix:effective_noise_model}, we go through details of our model of noisy quantum computation.
In Appendix~\ref{Appendix:optimized_algorithms}, we discuss principles of a quantum circuit optimization,
where one tailors the form of the effective Lindbladian.
In Appendix~\ref{sec:error_and_optimization_analysis}, we quantify the main error sources in our approach.
Finally, in Appendix~\ref{Appendix:bosons_from_fermions}, we discuss how to map a fermionic open-system model to
a spin-boson model.


\section{Open-system model}\label{sec:SimulatedModel}

While the plethora of physical phenomenon encountered in nature naturally correspond to a wide range of potential open system models,
we focus here on one concrete model of a specific form.
In particular, in its most general form, the open-system Hamiltonian we consider can be written
\begin{align}\label{eq:open_system_model1}
\hat H_0 &= \hat H_\textrm{S} + \hat H_\textrm{B} + \hat H_\textrm{C} \, ,
\end{align}
where $\hat H_\textrm{S}$ is the Hamiltonian of the system, $\hat H_\textrm{B}$ of the bath,
and $\hat H_\textrm{C}$ describes their coupling.

In this work, we explicitly consider problems that can be represented as a two-state system interacting with a bosonic bath~\cite{Spin_Boson_Rev}
by the so-called spin-boson model:
\begin{align}
\hat H_\textrm{S} &= -\frac{\hbar\Delta}{2}\hat\sigma_z \label{eq:spin_boson_model1} \, , \\
\hat H_\textrm{B} &= \sum_{k} \hbar\omega_{k} \hat b^\dagger_{k} \hat b_{k}\label{eq:spin_boson_model2} \, , \\
\hat H_\textrm{C} &= \hat\sigma_x\sum_k \frac{v_{k}}{2} \left(\hat b^\dagger_{k} + \hat b_{k}\right)  \, . \label{eq:spin_boson_model3}
\end{align}
The model system consists of a single spin which has an energy-level splitting $\hbar\Delta$.
The model bath consists of bosonic modes~$k$ with natural frequencies $\omega_{k}$.
The boson creation and annihilation operators satisfy $[\hat b_{k}, \hat b^\dagger_{l}]=\delta_{kl}$.
The coupling between the system and bath is here chosen to be transverse~[\onlinecite{Leppakangas2018_2}]
and occurs via the bath operator
\begin{align}
\hat X=\sum_{k} v_{k}(\hat b_{k}^\dagger + \hat b_{k}) \, .
\end{align}
The couplings $v_{k}$ are real numbers (possible phases were absorbed into the definitions of the boson operators).
A generalization to the multi-spin case 
is presented in Appendix~\ref{Appendix:multi_spin_systems}.

Since the bath Hamiltonian is non-interacting, its free-evolution statistics are Gaussian (time evolution according to $\hat H_\textrm{B}$).
This means that in the interaction picture,
a trace over the bath degrees of freedom results in an expression where the bath appears only via two-time correlation functions, whose Fourier Transform is the spectral function:
\begin{align}\label{eq:spectral_function_definition1}
S(\omega) = \int_{-\infty}^\infty dt e^{\textrm{i}\omega t} \left\langle \hat X(t) \hat X(0) \right\rangle_0 \, .
\end{align}
Here $\hat X(t) = e^{\textrm{i}t \hat H_\textrm{B}}\hat X e^{-\textrm{i}t \hat H_\textrm{B}}$ and
the index 0 refers to an average according to the free evolution of the bath.
The effect of the bath on the system is then fully determined by this function.

In thermal equilibrium we have:
\begin{align}\label{eq:spectral_function_definition2}
S(\omega) = 2\pi \sum_k \frac{ v_{k}^2\delta(\vert\omega\vert - \omega_{k})} {1-\exp\left( -\frac{\hbar\omega}{k_\textrm{B}T} \right)}\textrm{sign}(\omega) \, .
\end{align}
The temperature controls the symmetry (or lack-thereof)
between the positive (energy absorption by the bath)
and negative (energy emission by the bath) frequencies.
The functional form of the spectral function is important in the sense that it is not a constant
and therefore does not
correspond to just white noise. In particular, this implies that the bath has a memory, {\it i.e.}, it is non-Markovian.

The wide applicability of the spin-boson theory is based on the fact that the bath described by $\hat H_\textrm{B}$
does not necessarily need to consist microscopically of bosonic modes, but need only be effectively Gaussian.
An example of such bath is given in Appendix~\ref{Appendix:bosons_from_fermions}, where we derive a spin-boson model of a spin coupled to an electronic bath.


\section{Bath mapping}\label{sec:BathMapping}

\begin{figure*}
\includegraphics[width=1.55\columnwidth]{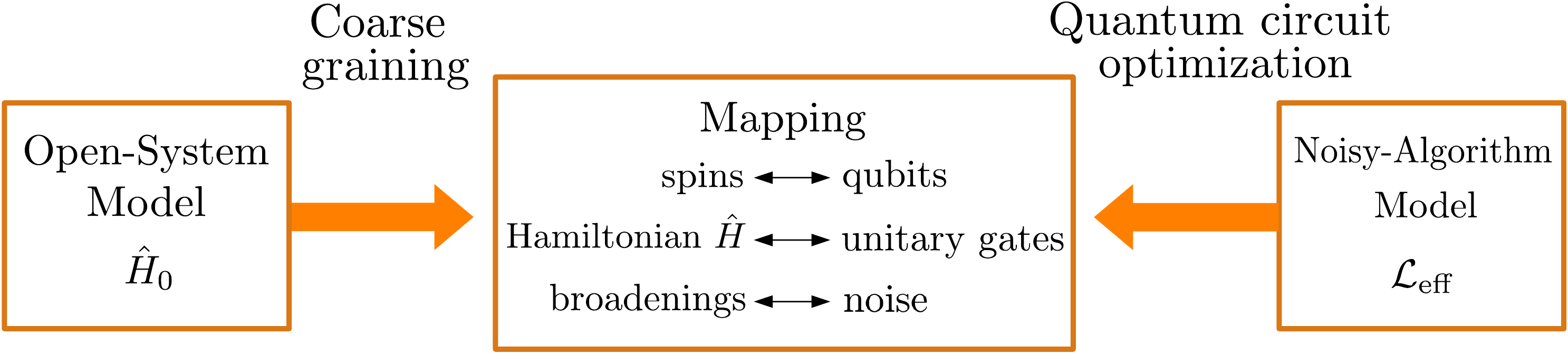}
\caption{Bath mapping, {\it i.e.}, mapping between an open-system model and a noisy-algorithm model.
As described in the main text,
the open-system model~$\hat H_0$ is first coarse grained to an auxiliary-spin model~$\hat H$ and additional spin broadening.
A quantum algorithm time evolves the qubits describing the spins according to~$\hat H$.
The combined effect of unitary gates and non-unitary qubit noise is described by an effective Lindbladian ${\cal L}_\textrm{eff}$.
The auxiliary-spin model broadening and the noise broadening according to ${\cal L}_\textrm{eff}$ are set equal by choosing the Trotter time step correctly.
A circuit optimization may be needed to have ${\cal L}_\textrm{eff}$ in a desired form.
}\label{fig:fitting_visualization3}
\end{figure*}

At the center of our approach is a mapping between an open-system model and a noisy-algorithm model.
The mapping is visualized in Fig.~\ref{fig:fitting_visualization3}.
The open-system model is first represented using a model that includes only spins, with Hamiltonian~$\hat H$, and additional spin broadenings.
The spins are represented by qubits on a quantum computer.
The noisy-algorithm model, on the other hand, describes the Trotter time evolution according to~$\hat H$
as a static Lindbladian ${\cal L}_\textrm{eff}$ operating on a time-evolved density matrix~\cite{Fratus2022}.
This indicates how noise added after each unitary gate appears as non-unitary processes in the simulated system,
see~Appendix~\ref{Appendix:effective_noise_model} and Ref.~\cite{Fratus2022}.
The spin broadenings and the noise described by ${\cal L}_\textrm{eff}$ are set equal by choosing the Trotter time step~$\tau$ correctly.
The full procedure is described in more detail below.

\subsection{Coarse graining}\label{sec:coarse_graining}
In this part of the mapping, we reduce the number of bath modes in the open-system model from infinite
to $n$~modes with broadening. These broad modes are now called auxiliary boson modes and later they will be mapped to the (noisy) bath qubits on the quantum computer.
As a practical matter, in this step we fit the target spectral function by $n$ Lorentzians.
We also write down a Lindblad master equation that is equivalent to this coarse-grained spectral function.

\subsubsection{Coarse graining when only the bath qubits are noisy}\label{sec:coarse_graining_scheme1}
\begin{figure}
\begin{center}
\includegraphics[width=1.0\columnwidth]{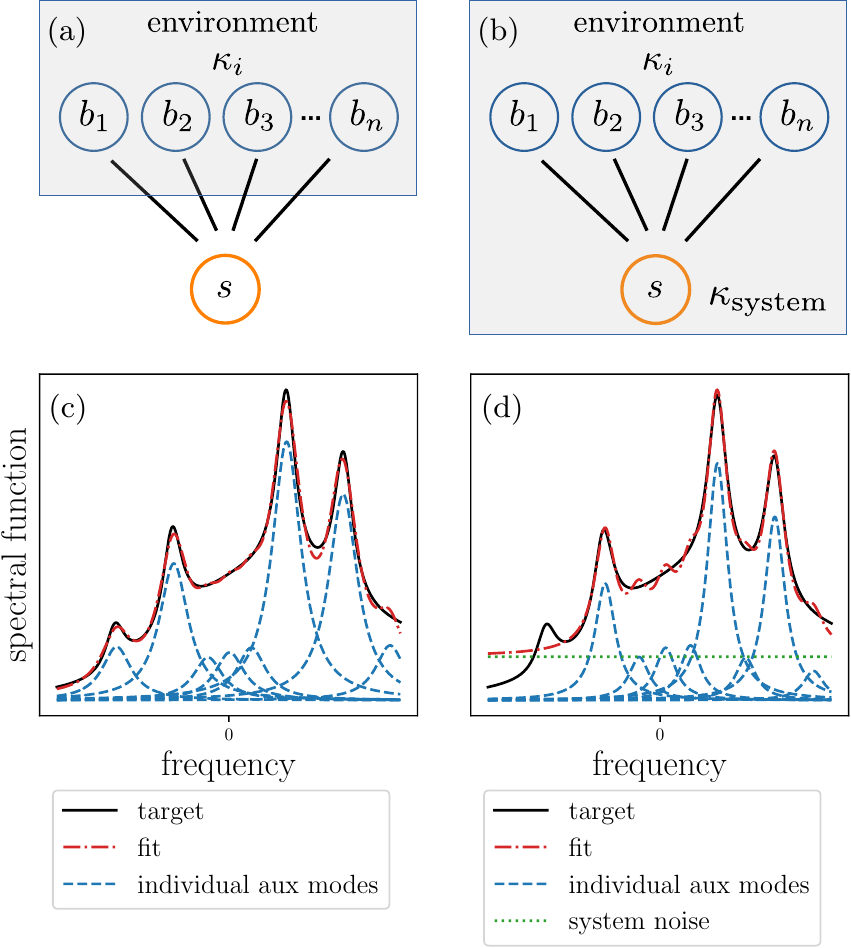}
\end{center}
\caption{
Fitting of a target spectral function in the two discussed bath-mapping schemes.
(a) The system spin~$s$ couples to $n$ auxiliary boson modes~$b_1,...,~b_n$.
The auxiliary modes further couple to an environment, which leads to mode broadenings~$\kappa_i$.
(b) In the second scheme, also the system spin couples to the environment, which
leads to a decoherence rate~$\kappa_\textrm{system}$.
(c) An example of fitting by eight auxiliary modes without system noise.
All modes are assumed to have the same broadening.
(d) An example of fitting the same spectral function but in the presence of system noise.
The system noise contributes via a constant shift of the fitting function,
leading to a different set of optimal auxiliary-mode parameters.
}\label{fig:fitting_visualization}
\end{figure}

In this coarse-graining scheme, the target spectral function, Eq.~(\ref{eq:spectral_function_definition2}), is fitted by $n$~Lorentzians,
\begin{align}\label{eq:coarse_grained_spectral_function}
S(\omega) &= \sum_{i=1}^n v_{i}^2 \frac{\kappa_{i}}{(\kappa_{i}/2)^2 + (\omega-\omega_{i})^2} \, .
\end{align}
Here we use the counting index~$i$ instead of~$k$, referring to the auxiliary modes.
The fitting is done by optimizing the peak areas $2\pi v^2_{i}$, frequencies~$\omega_{i}$, and broadenings~$\kappa_{i}$.
Later, the broadenings will be mapped to qubit noise.
Therefore, in the fitting,
the relative sizes of the broadenings need to be fixed so that they will be consistent
with the effective noise given by the noisy-algorithm model~${\cal L}_\textrm{eff}$ (Sec.~\ref{sec:noise_model_matching}).
The absolute size (a common prefactor) is a free fitting parameter,
since it will correspond to choosing the Trotter time step~$\tau$.
The coarse graining results shown in this paper are based on a least-squares fit,
whose details 
are presented in Appendix~\ref{Appendix:multi_spin_systems}.

The couplings and frequencies obtained in the fitting correspond to parameters of the Hamiltonian:
\begin{align}\label{eq:coarse_grained_Lindblad0}
\hat H &= \hat H_\textrm{S} + \hat\sigma_x\sum_{i=1}^{n} \frac{v_{i}}{2}\left(\hat b_{i}^\dagger + \hat b_{i}\right) + \sum_{i=1}^{n} \hbar \omega_{i} \hat b^\dagger_{i} \hat b_{i} \, ,
\end{align}
and the mode broadenings to damping rates in the Lindblad master equation:
\begin{align}\label{eq:coarse_grained_Lindblad}
\dot {\hat \rho} &= \frac{\textrm{i}}{\hbar}[\hat \rho, \hat H] + \sum_{i=1}^n\kappa_{i}\left( \hat b_{i} \hat \rho \hat b_{i}^\dagger - \frac{1}{2}\left\{\hat b_{i}^\dagger \hat b_{i} , \hat\rho \right\} \right) \, ,
\end{align}
where $\hat\rho$ is the density matrix of the spin and the auxiliary boson modes.

We note that
similar coarse-graining approaches have been presented earlier in the context of representing non-Markovian master equations
using pseudo modes~\cite{Arrigoni2013, Dorda2014, Tamascelli2018, Chen2019, Pleasance2020, Pleasance2021}.
Indeed, our work offers an optimal translation of such approaches to digital quantum simulation.
Similar coarse-graining approaches have also been presented in analog quantum simulation of spin-boson models~\cite{Mostame2012, Lemmer2018, Leppakangas2018_2, Potocnik2018}.
Also closely related are other exact numerical methods where one replaces continuous bosonic baths by a set of discrete modes,
such as the hierarchical equations of motion~\cite{Garraway1997, Ishizaki2005, Lambert2019}.

\subsubsection{Coarse graining when the system qubit is also noisy}\label{sec:coarse_graining_scheme2}
Our coarse graining can also account for decoherence of a qubit representing the (system) spin.
The decoherence of this system qubit contributes via a background rate in the spectral function.
Here, we proceed as before, except we instead optimize the spectral function
\begin{align} \label{eq:spectral_function_with_background}
S(\omega) &= \sum_{i=1}^{n} v_{i}^2\frac{\kappa_{i}}{(\kappa_{i}/2)^2 + (\omega-\omega_{i})^2} + 4\hbar^2\kappa_\textrm{system} \, .
\end{align}
The system noise appears via the rate $\kappa_\textrm{system}$, which however will not be an independent fitting variable.
In the fitting, the overall scale of the variables $\kappa$ is arbitrary,
but the relation between the bath mode broadenings $\kappa_i$ and the system noise $\kappa_\textrm{system}$ is fixed,
reflecting the corresponding hardware properties.
We then insert
\begin{align}\label{eq:noise_ratio}
\kappa_\textrm{system}= r \kappa \,,
\end{align}
where, for simplicity, we assume homogeneous bath broadening~$\kappa$.
If the system qubit noise matches the bath qubit noise, we have $r=1$, unless
a noise-symmetrization algorithm is used (Sec.~\ref{sec:SET}), where we have $r=1/2$.

The spectral function of Eq.~(\ref{eq:spectral_function_with_background})
corresponds to a time evolution according to the Lindblad master equation
\begin{align}\label{eq:coarse_grained_Lindblad2}
\dot {\hat \rho} &= \frac{\textrm{i}}{\hbar}[\hat \rho, \hat H]+ \sum_{i=1}^n\kappa_{i}\left( \hat b_{i} \hat \rho \hat b_{i}^\dagger - \frac{1}{2}\left\{\hat b_{i}^\dagger \hat b_{i} , \hat\rho \right\} \right) \nonumber\\
&+ \kappa_\textrm{system}\left( \hat\sigma_x \hat \rho \hat\sigma_x - \hat \rho\right) \, .
\end{align}
The difference to Eq.~(\ref{eq:coarse_grained_Lindblad}) is the term proportional to $\kappa_\textrm{system}$.
We see that the system collapse-operator in this term is $\hat\sigma_x$,
since the coupling to the bath is via this operator, see Eq.~(\ref{eq:coarse_grained_Lindblad0}).
It follows that the bath mapping in this scheme is exact
if the system-qubit noise operator (in the corresponding noisy-algorithm model) is also proportional to~$\hat\sigma_x$.
It should be noted that an exact mapping can be obtained also in the case of all qubits being subject to damping,
after a noise-symmetrization algorithm, as shown in Sec.~\ref{sec:SET}.
More generally, the bath mapping in such a scheme is most likely only an approximation,
when, for example, the system is subjected to depolarising noise,
but can work well for a weak coupling between the system and the bath.
The fitting in the two different schemes is illustrated in Fig.~\ref{fig:fitting_visualization}.

\subsection{Representing auxiliary bosons by auxiliary spins}\label{sec:from_bosons_to_spins}
\begin{figure*}
\includegraphics[width=1.4\columnwidth]{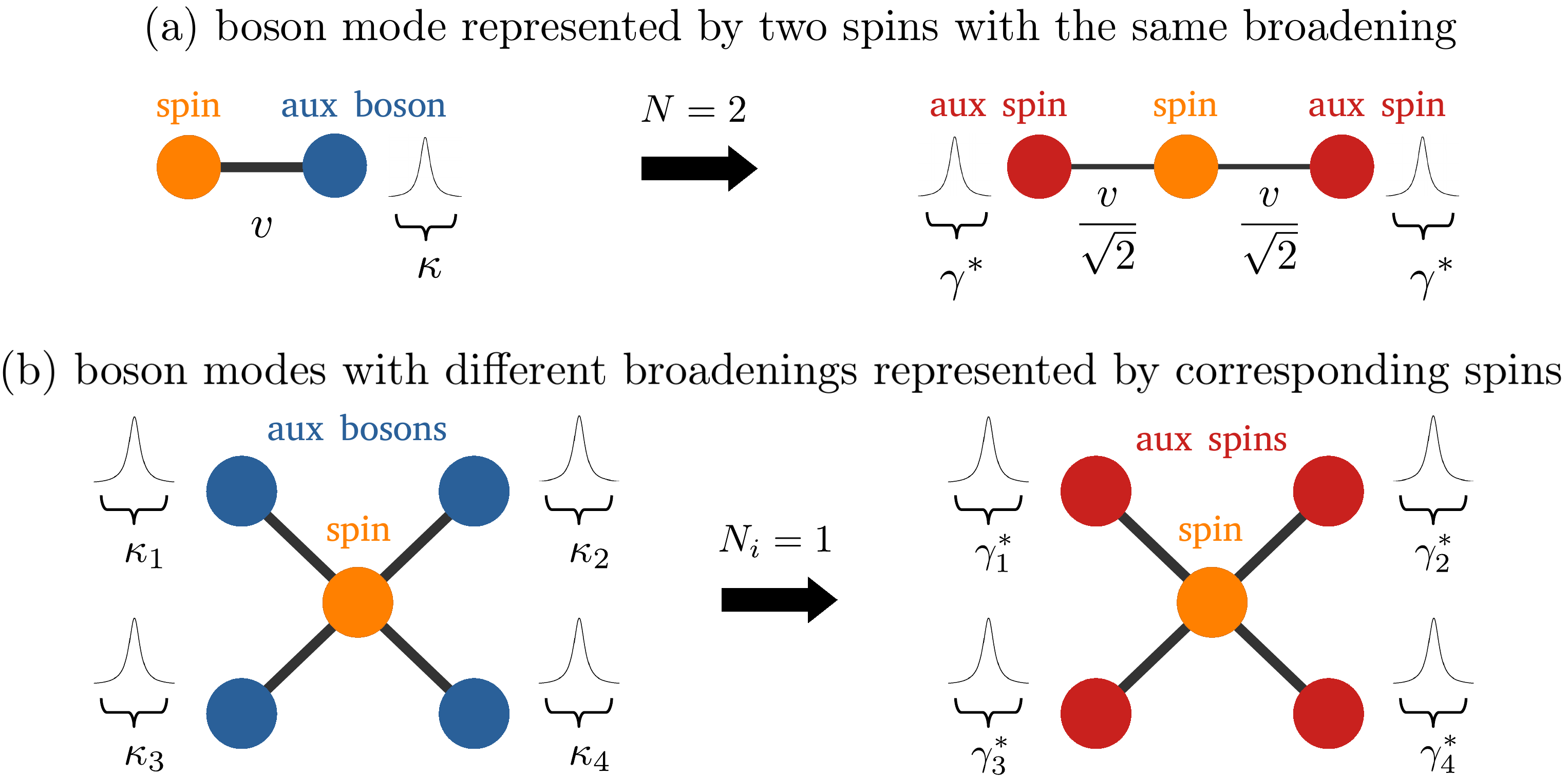}
\caption{Two possible approaches for representing auxiliary bosons by auxiliary spins (and finally by bath qubits).
(a) The coarse-grained spin-boson model includes one spin coupled to one sharp auxiliary boson mode.
The auxiliary boson mode is chosen to be represented by $N=2$ identical auxiliary spins. 
The new couplings are down-scaled by a factor $1/\sqrt{N}$ in order to keep the total spectral weight unchanged.
We have $\kappa=\gamma^*$, where the effective spin broadening is a sum of spin damping and dephasing, $\gamma^* = \gamma+2\Gamma$.
This mapping approach 
can be applied if the bath qubits representing the auxiliary spins have relatively homogeneous broadenings.
(b) An approach that can be applied when the bath qubits have significantly differing broadenings or when improvement of bath Gaussianity is not needed.
Here one imposes one-to-one correspondence between the boson modes and the auxiliary spins. 
}\label{fig:fitting_visualization2}
\end{figure*}

Here we write down a Lindblad master equation that includes only spins and is equivalent with the coarse-grained spin-boson model.
For this, we represent the derived auxiliary boson modes by auxiliary spins.
Common digital encodings~\cite{Sawaya2020} cannot be applied,
since they do not map damping of an arbitrary auxiliary spin to single-boson annihilation,
which is the key correspondence in our algorithm.
Instead, we replace bosonic energy operators $\hat b^\dagger \hat b$ by a sum of auxiliary spin operators $\hat\sigma^j_+\hat\sigma^j_-$
and bosonic coupling operators $\hat b^\dagger + \hat b$ by a sum of auxiliary spin operators $\hat\sigma_x^j$.
In other words, in the Hamiltonian~$\hat H$ we switch
\begin{align}
\hat b^\dagger_{i} \hat b_{i} &\rightarrow \sum_{j=1}^{N_{i}} \hat\sigma^{i,j}_+\hat\sigma^{i,j}_- \label{eq:spin_boson_operator_mapping1} \\
\left(\hat b_{i}^\dagger + \hat b_{i}\right)  &\rightarrow \frac{1}{\sqrt{N_{i}}} \sum_{j=1}^{N_{i}}\hat\sigma^{i,j}_x \label{eq:spin_boson_operator_mapping2} \, .
\end{align}
This mapping is based on the well-known property that collective spin operators can behave like bosonic operators
if their excitation numbers stay low, with an error ${\cal O}(1/N)$.
For a more detailed analysis of this point, see for example Ref.~\cite{Henschel2010}.
The coarse-grained model Hamiltonian, represented now by only spins, becomes
\begin{align}\label{eq:SpinSpinHamiltonian}
\hat H = \hat H_\textrm{S} + \hat\sigma_x \sum_{i=1}^n \frac{v_{i}}{2} \sum_{j=1}^{N_i}\frac{1}{\sqrt{N_i}} \hat\sigma^{i,j}_x + \sum_{i=1}^n \hbar\omega_{i} \sum_{j=1}^{N_i}\hat\sigma^{i,j}_+\hat\sigma^{i,j}_- \, .
\end{align}
The spectral function as seen by the system, Eq.~(\ref{eq:coarse_grained_spectral_function}) or~(\ref{eq:spectral_function_with_background}), keeps its form,
with the summation performed now over the corresponding auxiliary-spin parameters.


How the fitting and the replacement of the auxiliary bosons by auxiliary spins is done exactly, is
influenced by the open-system model as well as the hardware properties.
For example, for sharp model peaks in comparison to model couplings one may need $N_i>1$.
This choice implicitly assumes that the corresponding bath qubits have homogeneous decoherence rates, since the auxiliary spins are mapped later directly to bath qubits.
On the other hand, if significant differences exist in the bath-qubit decoherence rates,
or if the spectral function fitting is more important than the bath Gaussianity,
one should impose one-to-one correspondence between the auxiliary-boson modes and the auxiliary spins, {\it i.e.}, fix $N_i=1$.
These two choices are visualized in Fig.~\ref{fig:fitting_visualization2}.

It is very valuable for the bath mapping that
for auxiliary spins the spectral-peak broadening can be theoretically not only due to the damping rate~$\gamma$
but also due to the dephasing rate~$\Gamma$,
\begin{align} \label{eq:spin_broadening1}
\kappa_i &= \gamma_i + 2\Gamma_i \equiv \gamma^*_i\, .
\end{align}
This property is important since it allows one to do the full mapping procedure also when the bath qubits are subjected to significant dephasing.
The corresponding Lindbladian is (for simplicity considering here $N_i=1$)
\begin{align}\label{eq:effective_Lindblad}
\dot {\hat \rho} &= \frac{\textrm{i}}{\hbar} [\hat \rho, \hat H] + \sum_{i=1}^n \gamma_i\left( \hat\sigma_-^{i}\hat \rho\hat\sigma_+^{i} -\frac{1}{2}\left\{ \hat\sigma_+^{i}\hat\sigma_-^{i} , \hat \rho \right\} \right) \nonumber \\
& + \sum_{i=1}^n\frac{\Gamma_i}{2}\left( \hat\sigma_z^{i}\hat \rho\hat\sigma_z^{i} -\rho \right)  \, .
\end{align}
The procedure is otherwise the same as for bath qubits subjected to damping only,
except the connection between the broadenings of the bosonic bath and the hardware qubits is made via Eq.~(\ref{eq:spin_broadening1}).
We verify this connection by numerical simulations presented in Sec.~\ref{sec:spin_coupled_to_one_mode}.
[The possible system noise is added similarly as in Eq.~(\ref{eq:coarse_grained_Lindblad2}).]

It should be noted that broadening due to dephasing alone does not lead to Gaussian equilibrium statistics:
a finite damping rate $\gamma$ is needed.
This is because certain bath correlations that describe non-bosonic statistics decay in time according to $\gamma$ instead of $\gamma + 2\Gamma$.

\subsection{Matching spin broadening with qubit noise}\label{sec:noise_model_matching}

Here we describe how the Lindblad master equation operating only on spins can be implemented in digital quantum computing using noise.
The key part is the matching of the spin broadening with the qubit noise.
This step should be done consistently with the coarse graining (Sec.~\ref{sec:coarse_graining}):
variations in the noise of qubits
must appear as corresponding relative variations in the auxiliary-mode broadenings~$\kappa_i$.
This demand however does not lead to a self-consistency loop,
but rather to a unique determination of the variations of the auxiliary-mode broadenings.


In this work, we
assume that the qubit noise is predominantly damping and dephasing.
The form of the effective noise can however depend on the choice of gate decompositions,
which needs to be take into account when designing the quantum circuits.
This is detailed in Appendices~\ref{Appendix:effective_noise_model} and~\ref{Appendix:optimized_algorithms}.

In the considered case,
the noisy-algorithm model~(Appendix~\ref{Appendix:effective_noise_model} and Ref.~\cite{Fratus2022}) gives the effective spin damping and dephasing rates:
\begin{align}
\gamma_i &= \frac{D t_\textrm{gate} \bar\gamma_i}{\tau}\label{eq:assumed_noise_form1} \, , \\
\Gamma_i &=\frac{D t_\textrm{gate} \bar\Gamma_i}{\tau}\label{eq:assumed_noise_form2} \, ,
\end{align}
where $D$ is the depth of one Trotter-step circuit, $t_\textrm{gate}$ is the physical time needed to perform one gate
(assumed here to be a constant),
$\bar\gamma_i$ and $\bar\Gamma_i$ are the physical damping and dephasing rates of the qubit representing the auxiliary spin~$i$,
and $\tau$ is the chosen Trotter time step.
We assume here that the errors are similar for every gate 
and that they act also on qubits that are at rest (idling).
The potential variation in the widths $\gamma_i+2\Gamma_i$ needs to be accounted for in the coarse graining, as discussed above. 
The contribution from finite system qubit noise is similar. 

Eqs.~(\ref{eq:assumed_noise_form1}-\ref{eq:assumed_noise_form2}) with Eq.~(\ref{eq:spin_broadening1})
can be used to solve the Trotter time step~$\tau$, which was still a free parameter. 
This solution corresponds to matching the spin broadening with the qubit noise.
For simplicity, let us now assume that all bath qubits have homogeneous damping and dephasing rates,
and obtain
\begin{align}\label{eq:kappa_T_relation}
\tau &= \frac{D \epsilon}{\kappa} \, ,
\end{align}
where we have defined the gate error parameter
\begin{align}\label{eq:infidelity_main_text}
\epsilon \equiv t_\textrm{gate}(\bar\gamma+2\bar\Gamma)  \,.
\end{align}
An important observation 
is that the correct value of the Trotter time step~$\tau$ scales linearly with the gate error~$\epsilon$.
In terms of qubit decay time~$T_1=1/\bar\gamma$ and pure dephasing time~$T_\Phi=1/\bar\Gamma$,
the gate error is defined equivalently
\begin{align}\label{eq:infidelity_main_text2}
\epsilon = t_\textrm{gate}\left(\frac{1}{T_1} + \frac{2}{T_\Phi}\right) \,. 
\end{align}
In comparison to other error metrics in the literature~\cite{GoogleSupremacySI, Abad2022},
the contribution from pure dephasing is here doubled.
This is because we do not average over the Bloch sphere, but instead look at noise in the Z-basis. When all qubits are subject to damping only,
the corresponding one-qubit Pauli (or average) error is~$\epsilon/2$~($\epsilon/3$) and the two-qubit error is~$\epsilon$~($4\epsilon/5$).

Since all terms of the normalized Hamiltonian $\hat H/\kappa$ are fixed in the coarse graining,
Eq.~(\ref{eq:kappa_T_relation}) gives the angles of the unitary gates
$\exp\left(-\textrm{i}\hat H \tau/\hbar\right)^m=\exp\left(-\textrm{i} D\epsilon \hat H/\hbar\kappa\right)^m$, {\it i.e.},
the angles of the unitary gates in the Trotterization of the time-evolution operator over simulation time $t=m\tau$.
This completes the mapping.


\begin{figure*} 
\includegraphics[width=2\columnwidth]{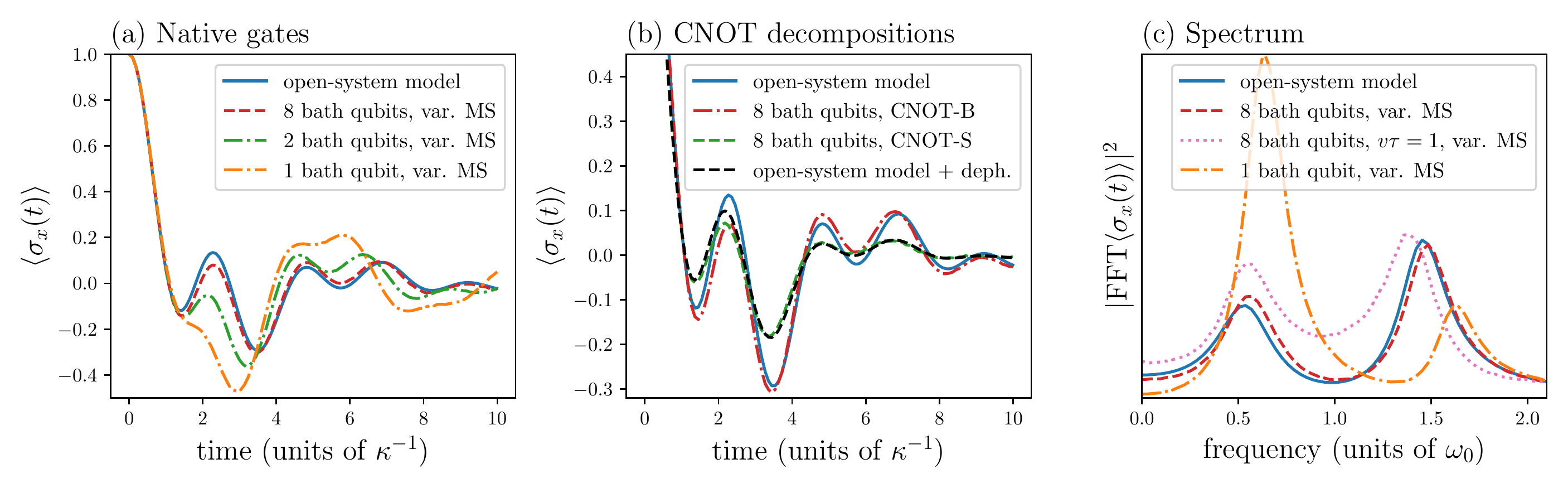}
\caption{Open-system dynamics of a spin coupled to a broad bosonic mode solved using the quantum algorithm
and a comparison to the numerically exact solution (open-system model).
We plot the time evolution of $\left\langle \hat\sigma_x(t) \right\rangle$
when the system spin is initially excited to +1 eigenstate of $\hat\sigma_x$ and the bath mode is at ground.
When time evolving on the quantum computer, the bath qubits are subject to damping and dephasing.
We consider model parameters $v=\omega_0=2\kappa$ and in (a-b) $\Delta = 0.9\omega_0$, in (c) $\Delta=\omega_0$.
(a) An agreement with the open-system model is achieved when multiple qubits are used to represent
the significantly populated bosonic mode. This improves the Gaussianity of the quantum-algorithm bath.
We consider here a native (variable MS) gate decomposition and a gate error of
$1$\%, $3$\%, and $5$\% for $N=8$, 2, 1, respectively.
(b) When $N=8$, the result for the non-native decomposition CNOT-B (bath as control) is practically the same as for the variable MS,
but the result for CNOT-S (system as control) differs due to the emergence of effective system-spin dephasing.
We consider here a gate error of~$0.36$\%.
(c) The spectrum of $\left\langle \hat\sigma_x(t) \right\rangle$ is characterized by a peak splitting by coupling $v$.
Also plotted is the result for a large Trotter step $v\tau=1.0$, corresponding to a gate error of~$5.6$\%.
}\label{fig:one_broad_mode}
\end{figure*}

\section{Examples}\label{sec:examples}

In this section we present examples of solving open-system dynamics using our quantum algorithm.
We solve dynamics for three different models
and study the quality 
by comparing to exact or approximate solutions derived in the literature.
In Sec.~\ref{sec:spin_coupled_to_one_mode}, we study the dynamics of a spin coupled ultra-strongly to a resonance mode with broadening.
In Sec.~\ref{sec:spin_coupled_to_ohmic_bath}, we study a spin coupled to an ohmic bath.
In these examples we assume that the quantum computer has a noiseless system qubit.
In Sec.~\ref{sec:SET}, we study the steady state and relaxation dynamics of strongly interacting electrons hopping between an island and leads.
In this example, the system qubit is allowed to be noisy too.
The numerical results shown for the Trotter time evolution
were obtained using {\verb|qoqo|}~\cite{qoqo} with \verb|QuEST|~\cite{quest} as the simulator backend,
by solving the time evolution of the density-matrix of the qubits.
We assume all-to-all qubit-connectivity. However,
an efficient algorithm for a nearest-neighbor connectivity also exists and is presented in Appendix~\ref{Appendix:optimized_algorithms},
leading to at worst a~$50$\% increase in qubit number.

\subsection{Spin coupled ultra-strongly to a resonance mode with broadening}\label{sec:spin_coupled_to_one_mode}

Here we study the case of a bath with a resonance frequency.
We assume a Lorentzian spectral function, which can be coarse grained exactly by single auxiliary boson mode (sec.~\ref{sec:coarse_graining}).
The coarse-grained open-system model has the Hamiltonian:
\begin{align}
\hat H = -\frac{\hbar\Delta}{2}\hat\sigma_z + \frac{v}{2}\hat\sigma_x(\hat b^\dagger + \hat b) + \hbar\omega_0 \hat b^\dagger \hat b\, .
\end{align}
We consider the case of ultra-strong coupling, $v=\omega_0\sim\Delta$, and $\kappa=v/2$, which leads to clear non-Markovian system-bath dynamics.

In the next step of the bath mapping, the auxiliary boson mode is represented by~$N$ auxiliary spins (Sec.~\ref{sec:from_bosons_to_spins}).
On the quantum computer, auxiliary spins correspond to bath qubits. We assume all bath qubits have the same damping and dephasing rates,
$\bar\gamma$ and $\bar\Gamma$. We assume a significant dephasing by setting $\bar\Gamma = \bar\gamma/2$ ($T_\Phi = 2 T_1$).
The time evolution is performed according to a Trotter expansion of the time-evolution operator $\exp\left(-\textrm{i}\hat H \tau/\hbar\right)^m$,
decomposed into different two-qubit gates (see below), with total simulated time $t=m\tau$.
The Trotter time step~$\tau$ is chosen so that the auxiliary-spin broadening~$\kappa$ matches to the effective qubit noise, see Sec.~\ref{sec:noise_model_matching}.
This value is given by Eq.~(\ref{eq:kappa_T_relation}).
The circuit depths $D$ are listed in Table~\ref{table:circuit_depths}.

In Fig.~\ref{fig:one_broad_mode}, we show the time evolution of the expectation value $\left\langle \hat\sigma_x(t) \right\rangle$
after the system spin is excited to +1 eigenstate of $\hat\sigma_x$. The bath mode is initially at ground.
In Fig.~\ref{fig:one_broad_mode}(a), we compare the solutions obtained
when using different number of bath qubits~$N$ (auxiliary spins) to the numerically exact solution of the bosonic open-system model.
The time evolution on the quantum computer is implemented using variable MS two-qubit gates,
which is a native gate of the time-evolved Hamiltonian, Eq.~(\ref{eq:SpinSpinHamiltonian}).
We see that the quantum-algorithm solution approaches the correct one when the bath-qubit number~$N$ is increased.
This demonstrates that the algorithm works for bath qubits subjected to damping and dephasing,
and that the auxiliary-spin bath behaves like one bosonic mode after multiple spins are used to represent it.
The latter result is due to the ultra-strong coupling between the system and bath.
For weak system-bath couplings, $N=1$ would be adequate.
The Trotter time steps are here chosen to be $v\tau= 0.18, 0.18, 0.2$, for $N=8$, 2, 1, respectively.
They are chosen to be relatively large,
yet still small enough that they lead to negligible Trotter error on the visual scale of the plot.
These choices fix the assumed gate errors $\epsilon$ to~$1$\%, $3$\%, and $5$\%, respectively.

%

In Fig.~\ref{fig:one_broad_mode}(b), we study non-native gate decompositions. 
We show the results for two CNOT-decompositions with restricting to the case of $N=8$ bath qubits.
The optimal decomposition consists here of a CNOT, a small-angle X-rotation, and another CNOT.
There are still two versions of this circuit: either the bath qubits being the control qubits (CNOT-B) or the system qubit being the control qubit (CNOT-S).
The results differ since the CNOT operation as well as the noise properties are not symmetric with interchanging the system and bath.
We find that the result for the non-native decomposition CNOT-B is the same as for the native variable MS result (indistinguishable on the scale of the plot).
This is because the bath qubits are never operated on by the large-angle gates and therefore the noise does not get transformed into another form.
The results for optimal control-Z decompositions, shown in Appendix~\ref{Appendix:optimized_algorithms}, are similar.
However, in the case of CNOT-S, we find a noticeable difference to the correct open-system model result, which implies changes in~${\cal L}_\textrm{eff}$.
It can be derived analytically (Appendix~\ref{Appendix:effective_noise_model}),
that here its approximate form includes system-qubit dephasing, which emerges
even though the physical system-qubit is noiseless, the origin being the X-gates operated on the noisy bath qubits.
This interpretation is verified by additional numerical simulations,
in which we add system-spin dephasing to the open-system model, see Fig.~\ref{fig:one_broad_mode}(b).
We then find that the non-native decomposition (CNOT-S) is not perfect,
but still reproduces similar dynamics, with faster decay of coherence.

In Fig.~\ref{fig:one_broad_mode}(c), we study the spectrum of the obtained time evolutions.
We plot here the absolute square of the Fast Fourier Transform (FFT) of $\left\langle \hat\sigma_x(t) \right\rangle$.
The correct open-system model result shows a splitting of a resonance
located originally at $\omega_0$ to frequencies $\omega_0\pm v/2$.
This is reproduced well by the quantum algorithm with the bath-qubit number $N=8$ and the gate error of $1$\% ($v\tau=0.18$).
We also show the result for a very large Trotter step, $v\tau=1.0$ and $N=8$, corresponding to a gate error as large as~$\epsilon=5.6$\%.
We find that this result, with non-negligible Trotter error and sampling error, 
is clearly closer to the correct result than the one with $N=1$ ($v\tau=0.18$, gate error of $4.5$\%), which has significant bath-Gaussianity error.
This result is specific for the case of ultra-strong coupling,
where the bath excitation number is large and needs to be accounted for by large $N$, with the possible cost of increased Trotter error.
An error tradeoff in the quantum algorithm is discussed more in Appendix~\ref{sec:error_and_optimization_analysis}.

\subsection{Spin coupled to an ohmic bath}\label{sec:spin_coupled_to_ohmic_bath}

\begin{figure}
\includegraphics[width=0.7\columnwidth]{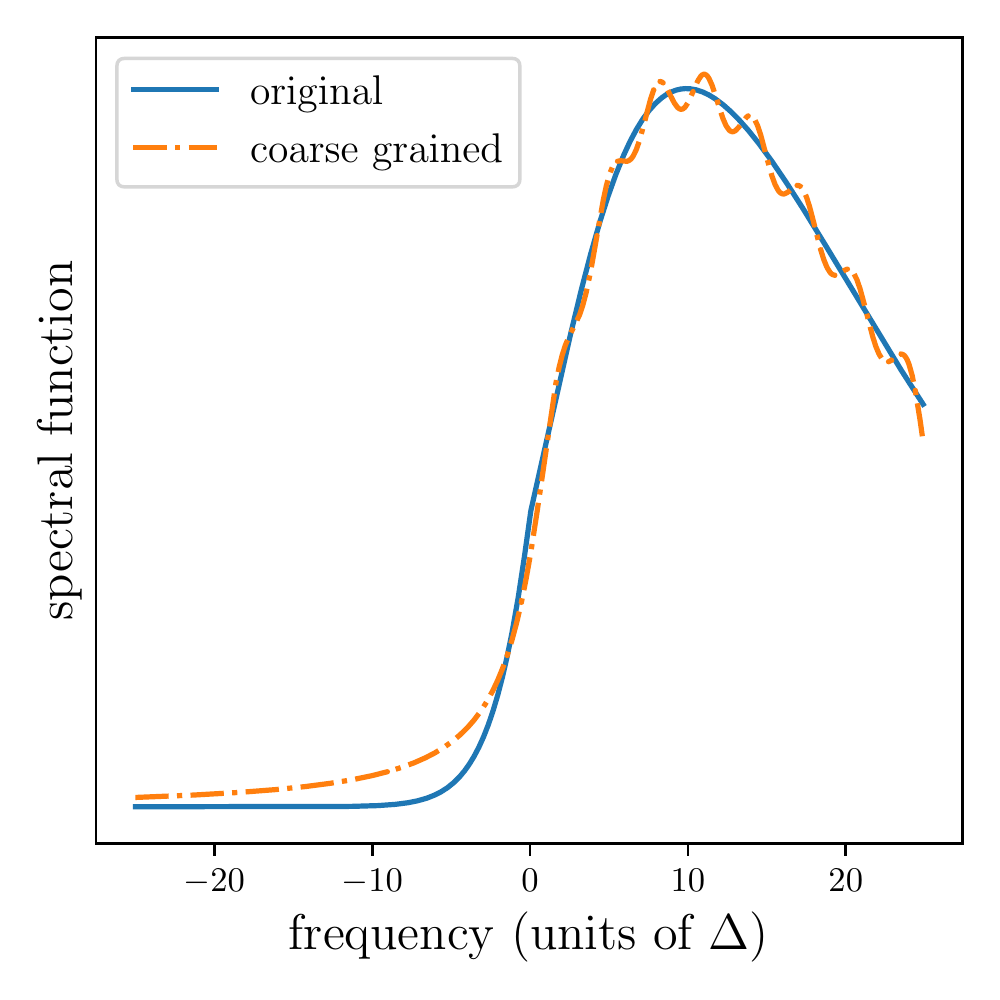}
\caption{Fitting of an ohmic spectral function with exponential cutoff by eight broad (auxiliary) boson-modes.
The fitting is done with a constriction of auxiliary modes having identical broadenings~$\kappa_i=\kappa$.
}\label{fig:ohmic_coarse_graining}
\end{figure}

\begin{figure*}
\includegraphics[width=2\columnwidth]{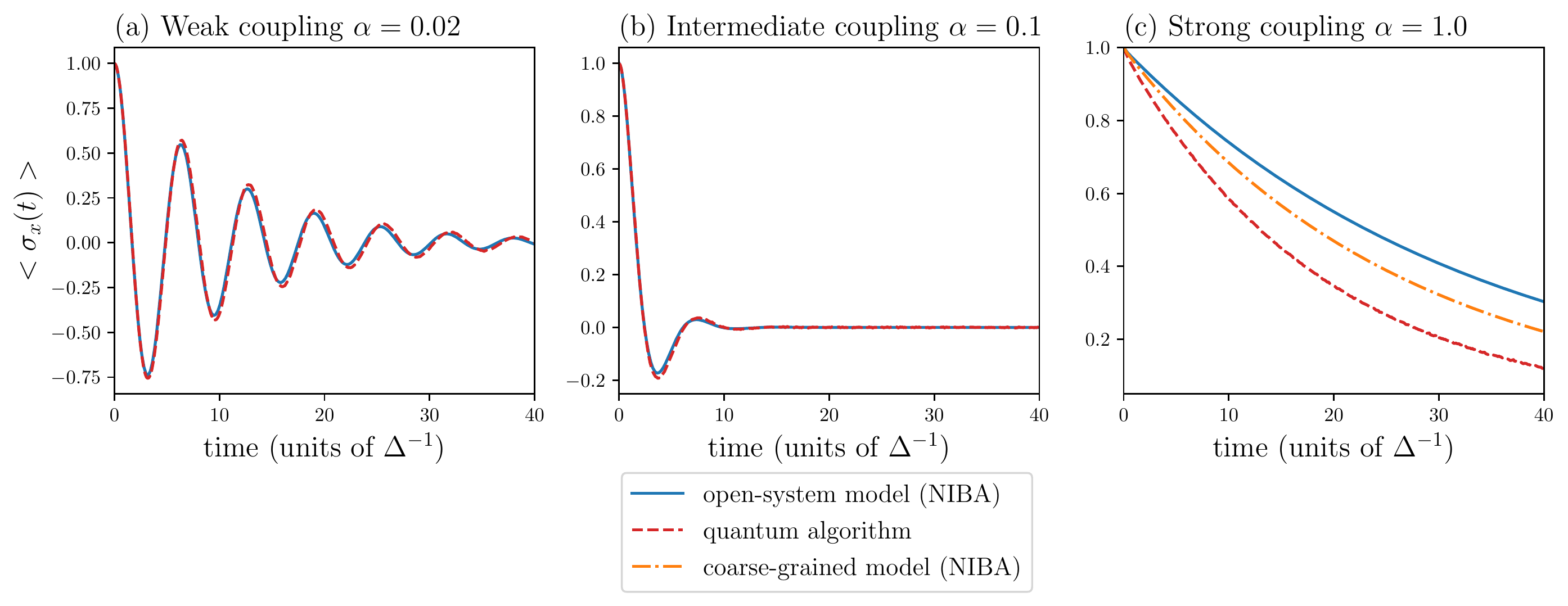}
\caption{Open-system dynamics of a spin coupled to an ohmic bath solved using the quantum algorithm
and a comparison to the numerical solution obtained using the NIBA (Non-Interacting Blip Approximation).
We use a variable MS decomposition and a gate error of~$5$\%.
In the quantum algorithm, the ohmic bath is represented by eight auxiliary boson-modes with a coarse-grained spectral function as shown Fig.~\ref{fig:ohmic_coarse_graining}.
Each of the boson modes is represented by one bath qubit.
We show the NIBA result also for the coarse-grained spectral function in the case of strong coupling $\alpha=1$
(while for $\alpha=0.02$ and $\alpha=0.1$ the differences are negligible).
We find that the famous transition from a weak damping to a slow decay with increasing $\alpha$ is reproduced well by the quantum algorithm.
}\label{fig:ohmic_spectrum}
\end{figure*}

The case of an ohmic bath emerges in many areas of physics 
and highlights how the system-bath dynamics can drastically change with the system-bath interaction strength.
We now show that this behavior is reproduced correctly by the presented quantum algorithm.

An ohmic spectral function has the form,
\begin{align}\label{eq:ohmic_spectral_function}
S(\omega) &= \frac{4\pi\hbar^2\alpha\omega}{1 - \exp\left(-\frac{\hbar\omega} {k_\textrm{B} T}\right) } \, ,
\end{align}
where the interaction strength is characterized by the so-called Kondo parameter $\alpha$ as well as the temperature~$T$.
We choose $k_\textrm{B}T/\hbar\Delta = 1.5$, where $\hbar\Delta$ is the spin splitting.
We also introduce a cutoff function $\exp\left(-\vert \omega\vert/\omega_\textrm{c}\right)$ with
$\omega_\textrm{c}/\Delta= 10$.
The coarse graining is done by eight auxiliary boson-modes with a constriction to identical broadenings~$\kappa_i=\kappa$
and is shown in Fig.~\ref{fig:ohmic_coarse_graining}.

In the quantum algorithm,
we choose to represent each of the auxiliary boson modes by one auxiliary spin.
When time evolving on the quantum computer, we assume that the corresponding bath qubits are subjected only to damping.
The quantum algorithm is of the same form as for the resonant bath considered in Sec.~\ref{sec:spin_coupled_to_one_mode},
with the difference that here the auxiliary spin parameters are not identical.
We consider the case of variable MS decomposition and a gate error~$\epsilon=5$\% ($\Delta \tau=0.11$).
The conclusions made for the use of different gate decompositions in Sec.~\ref{sec:spin_coupled_to_one_mode} are valid here as well.

In Fig.~\ref{fig:ohmic_spectrum}, we show the time evolution of $\left\langle \hat\sigma_x(t) \right\rangle$
when the system-spin is again set initially to the +1 eigenstate of $\hat\sigma_x$ and when the bath is initially at equilibrium.
We show the resulting dissipative dynamics for weak, intermediate, and strong coupling limits, corresponding to $\alpha=0.02, 0.1, 1.0$.
We compare the quantum algorithm results to the so-called NIBA (non-interacting blip approximation) calculation of the same dynamics~\cite{Spin_Boson_Rev, Magazzu2017}.
We perform this calculation for the original ohmic spectral function as well as for the coarse-grained spectral function. 
Since a noticeable difference appeared only for strong couplings, the latter result is shown only for $\alpha=1.0$.
We find that the celebrated transition, from a weak-to-strong damping and finally to a slow decay,
is reproduced well by the quantum algorithm.
We find a good agreement with the NIBA even though each of the auxiliary boson modes is represented by only one auxiliary spin.
A difference to the NIBA emerges only for large $\alpha$.
This difference is due to a combination of imperfect fitting (see the result for the NIBA with the coarse-grained bath)
and elevated bath-qubit populations at low-frequencies. 

\subsection{Electronic transport models}\label{sec:SET}

\begin{figure*}
\includegraphics[width=2\columnwidth]{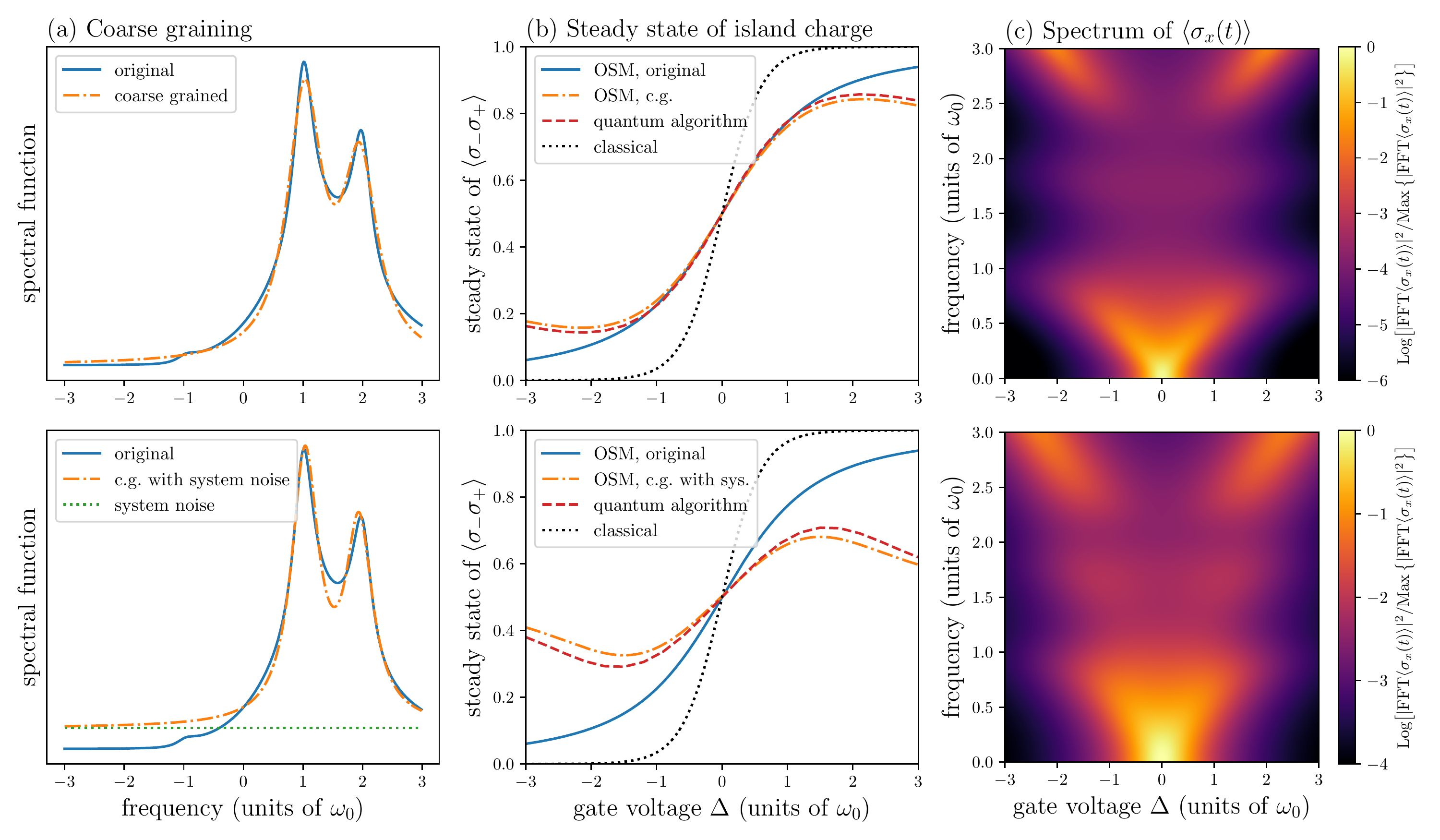}
\caption{The steady-state and relaxation dynamics of a single-electron transistor island-charge, solved using two quantum algorithms
(above, without system noise, below, with system noise),
and a comparison to a numerical solution obtained by a method presented in Ref.~\cite{SchoellerSchon1994} (referred as the open-system model, OSM).
We set the source-drain bias to zero and vary the gate voltage~$\Delta$.
Both algorithms are based on variable MS gate decomposition and a gate error of~$1$\%.
(a) Fitting of the resonant spectral function by two modes with identical broadenings (above) and
by two modes with identical broadenings and system noise (below).
(b) A comparison between the solutions for the steady-state island charge.
We calculate the OSM solution for the original spectral function as well as for the coarse-grained spectral function.
A difference appears at high gate voltages due to an increased effective temperature of the coarse-grained bath at high energies.
In the presence of the system-qubit noise (below),
a difference to the original open-system model result appears at all gate voltages due to an increased effective temperature at all energies.
The quantum algorithm result and the OSM result for the coarse-grained bath however stay similar.
(c) The spectrum of the charge relaxation dynamics (log-scale) when initially setting the island charge to +1 eigenstate of $\hat\sigma_x$.
A peak tripling appears in the vicinity of the bath resonances. 
When solved with the algorithm with system noise (below), additional broadening appears but the key characteristics remain.
}\label{fig:SET}
\end{figure*}

In the third example, we time evolve open-system dynamics according to a generalized spin-boson model that can describe electronic transport,
for example, across metallic islands or quantum dots.
This model is also interesting since it allows for mapping system-qubit damping to the coarse-grained spectral function exactly.
The Hamiltonian~$\hat H_0=\hat H_\textrm{S}+\hat H_\textrm{B}+\hat H_\textrm{C}$ has the form
\begin{align}
\hat H_\textrm{S} &= -\frac{\hbar\Delta}{2}\hat\sigma_z \, ,  \nonumber \\
\hat H_\textrm{B} &= \sum_{k} \hbar\omega_{k} \hat a^\dagger_{k} \hat a_{k} + \sum_{k} \hbar\omega_{k} \hat b^\dagger_{k} \hat b_{k} \, , \label{eq:SET_bath} \\
\hat H_\textrm{C} &= \hat\sigma_x\sum_k \frac{v_{k}}{2} \left(\hat a^\dagger_{k} + \hat a_{k}\right) + \hat\sigma_y\sum_k \frac{v_{k}}{2} \left(\hat b^\dagger_{k} + \hat b_{k}\right) \label{eq:SET_coupling} \, .
\end{align}
The model bath has two sets of bosonic modes,
described by the bosonic operators $\hat a$ and $\hat b$, coupled to the system via operators $\hat\sigma_x$ and $\hat\sigma_y$.
The spectral functions of these individual baths are defined similarly as before, Eq.~(\ref{eq:spectral_function_definition2}),
and are here identical. 
An example of an open quantum system that can be modeled with this Hamiltonian is a single-electron transistor (Appendix~\ref{Appendix:bosons_from_fermions}).
Here, for strongly interacting electrons only two island-charge states are relevant,
and these are mapped to the two states of the introduced system-spin.
The gate voltage of the transistor translates into the energy-level splitting $\Delta$.
The source-drain bias is assumed to be here zero.

We construct two quantum algorithms for solving this problem. The first one is designed for a quantum hardware
with a noiseless system-qubit and bath qubits that are subjected to damping noise,
and the second algorithm for all qubits being noisy.
Furthermore, we consider a structured spectral function~$S(\omega)$, which is ohmic at low frequencies, has resonances at $\omega_0$ and $2\omega_0$,
and is at relatively low temperature $k_\textrm{B}T=0.3\hbar\omega_0$ (Appendix~\ref{Appendix:bosons_from_fermions}).
This can describe transmission lines with resonances. The system-bath coupling is of similar magnitude as the resonance mode broadening.
The spectral function fitting corresponding to the two algorithms is shown in Fig.~\ref{fig:SET}(a).

In the algorithms, each auxiliary boson mode is represented by one bath qubit.
We have in total four bath qubits (two independent baths).
On hardware, this could be realized by a system qubit with four neighboring bath qubits.
We use a variable~MS decomposition and a gate error of~$1$\%.
In the case of noisy system qubit, X-gates are introduced between the original Trotter-step circuits in order to transform damping to
a sum of X-noise and Y-noise, which allows it to be mapped to the spectral function, see Appendix~\ref{Appendix:optimized_algorithms}.
We have a background rate factor $r=1/2$, see Sec.~\ref{sec:coarse_graining_scheme2}.

In Fig.~\ref{fig:SET}(b), we show the obtained steady-state values for the island charge
$\left\langle \hat\sigma_-\hat\sigma_+ \right\rangle $ as a function of the gate voltage $\Delta$. 
We also do a comparison to an open-system model solution
obtained by applying the numerical method developed in Ref.~\cite{SchoellerSchon1994}.
We again apply the reference method to the original spectral function as well as to the coarse-grained spectral function.
The shown classical solution is the result in the limit $\alpha\rightarrow 0$ and corresponds to the Fermi function.
In the case of a noiseless system qubit (above plot),
we find a good agreement between the quantum algorithm result and the open-system model solution.
In particular, all different results (except the classical one) are practically identical when the bias $\Delta/\hbar$ is at the well-fitted spectral-function region.
For larger $\vert\Delta/\hbar\vert$, differences appear due to imperfect coarse graining.
However, importantly, the results for the same spectral functions stay practically identical.
The coarse-grained spectral function can be interpreted as describing a non-equilibrium bath, with elevated temperature at higher energies.
When all qubits are subject to damping (bottom plot),
the steady-state island charge is now different for all gate voltages, when comparing to the original open-system model solution.
However, the results for the same spectral functions stay again almost identical.
This confirms that the system-qubit noise can indeed be mapped to the spectral function of the open-system model.

In Fig.~\ref{fig:SET}(c), we study the dynamical behavior of the island charge, when 
exciting the island charge initially to symmetric superposition, {\it i.e.}, to $+1$ eigenstate of $\sigma_x$.
We present the result for the spectrum of $\langle \hat\sigma_x(t)\rangle$.
In the case of the noiseless system qubit (above plot),
we find that far from the resonances $\Delta=\omega_0$ or $\Delta=2\omega_0$,
a narrow peak appears at frequency $\omega\approx\Delta$.
This corresponds to coherent phase oscillations between the charge states with exponential decay
of the coherence. The system-bath dynamics are Markovian.
Closer to the bath resonances, the peak shows strong broadening as well as splitting into a triplet.
We find that in this region the underlying system-bath dynamics is non-Markovian,
which occurs due to the strong system-bath coupling. 
We also see that the same key characteristics appear when solving the problem with the algorithm with system noise (bottom plot).
This highlights that dynamics of open quantum systems with a structured bath
can be solved with the presented quantum algorithm also in the presence of strong system-qubit noise.

\begin{table*}
\begin{center}
\begin{tabular}{ c | c | c | c | c | c | c }
\multicolumn{1}{c|}{} & \multicolumn{2}{c|}{Example A} & \multicolumn{2}{c|}{Example B} & \multicolumn{2}{c}{Example C} \\ 
\cline{2-7}
\multicolumn{1}{c|}{Decomposition} & Depth $D$ & gate error $\epsilon$~(\%) & Depth $D$ & gate error $\epsilon$~(\%) & Depth $D$  & gate error $\epsilon$~(\%) \\ 
\hline
\hline
Variable MS & $1+n_\textrm{q}$ & $1$ & $1+n_\textrm{q}$ & $5$ & $1 + 2n_\textrm{q}$ & $1$ \\ 
\hline
Variable iSWAP & $1+4n_\textrm{q}$ & $0.27$ & $1+4n_\textrm{q}$ & $1.4$ & $1 + 2n_\textrm{q}$ & $1$ \\ 
\hline
CNOT & $1+3n_\textrm{q}$ & $0.36$ & $1+3n_\textrm{q}$ & $1.8$ & $1 + 4n_\textrm{q}$ & $0.53$ \\ 
\hline
Control-Z & $1+5n_\textrm{q}$ & $0.2$ & $1+5n_\textrm{q}$ & $1.1$ & $1 + 6n_\textrm{q}$ & $0.36$ \\ 
\hline
\end{tabular}
\end{center}
\caption{Circuit depths and gate errors in the three examples presented in Sec.~\ref{sec:examples} when realized using different two-qubit gate decompositions.
The gate error is defined in Eqs.~(\ref{eq:infidelity_main_text}-\ref{eq:infidelity_main_text2}).
Access to lower gate error~$\epsilon$ allows the use of a smaller Trotter step (reduce the Trotter error) and/or a larger total number of bath qubits~$n_\textrm{q}$
(reduce the coarse graining and/or the bath Gaussianity error).
In the three examples we have $n_\textrm{q}=8,8,4$ respectively.
We assume a system-to-all-bath device connectivity.
For a nearest-neighbor connectivity, a system-bath swap-network can be applied (Appendix~\ref{Appendix:optimized_algorithms}),
leading to at worst a~$50$\% increase in qubit number and a $(n_\textrm{q}/2-1)n_\textrm{SWAP}$ increase in circuit depth,
where $n_\textrm{SWAP}$ is the number of gates needed to perform the SWAP operation and we assume an even $n_\textrm{q}$.
}\label{table:circuit_depths}
\end{table*}


\section{Discussion}\label{sec:discussion}
In this work we have developed a new framework for noise-utilizing quantum simulations.
The presented quantum algorithms map typical noise in digital quantum computing to spectral function properties 
of open system models.
At the center of this approach is the noisy-algorithm model,
which describes the effect of noise in the form of a static Lindblad master equation operating on the time-evolved
density matrix~\cite{Fratus2022}.

The form of this static Lindbladian depends on the used circuit decomposition.
In the examples we have considered, optimal circuit decompositions were based on native gates,
such as the small-angle single-qubit rotations, variable MS two-qubit gates (XX interaction),
or variable iSWAP two-qubit gates (XX + YY interaction).
We also found that non-native gate decompositions can work as well,
but may also lead to alternations to the actually time-evolved open-system model.
The examples considered also showed that the algorithm can perform very well even with gate errors as high as~1\%.
The performance however improves with decreasing gate error,
as this allows for the implementation of longer circuits, e.g., using smaller Trotter time step.
It is thus beneficial to have as low of a gate error as possible.
The central parameters used in the examples are summarized in Table~\ref{table:circuit_depths}.

In the shown examples we considered single-spin systems.
It should be emphasized that the quantum algorithm can be applied also to multi-spin systems, such as exciton-transport models~\cite{Ishizaki2012}.
Actually, the algorithm is most efficient when time evolving a model with multiple system spins,
due possibility to more efficient gate parallelization. More specifically, for a fixed number of bath modes,
system spins can be added without additional cost ({\it e.g.}, increase in the Trotter time step)
when exploiting all-to-all connectivity or the system-bath swap-network presented in Appendix~\ref{Appendix:optimized_algorithms}.

The derivation of the quantum algorithm assumed the use of qubits with intrinsic noise that can be described using a Lindbladian formalism.
However, we note that the approach is also directly applicable to more general quantum elements and noise whose effect is rather of Bloch-Redfield form.
Furthermore, the general principles of modifying the effective noise could also be applied to general non-Markovian noise.
Utilization of non-Markovian noise in quantum computing has been addressed recently in Ref.~\cite{Smart2022}.

The focus in this article has been on utilizing intrinsic bath-qubit noise. 
However, the bath qubits can also be replaced by resonators or phonon modes on hardware~\cite{Mostame2012, Lemmer2018, Leppakangas2018_2, Potocnik2018}
with engineered damping, which can offer more efficient simulations for large boson numbers.
It may also be possible to replace the system qubits by error-corrected qubits,
while having a large number of physical qubits as the bath.

Finally, the quantum algorithm and conclusions presented in this paper are not restricted to simulating the spin-boson model,
but are applicable also for more general open-system models.

\section*{Acknowledgments}
This work was supported by the
German Federal Ministry of Education and Research, through PhoQuant (13N16107), and QSolid (13N16155),
and by 
the German Federal Ministry of Economic Affairs and Climate Action, through the PlanQK project (01MK20005H).
This work was also supported by the European Union’s Horizon 2020 program number 899561, AVaQus.

\appendix

\section{Generalization to multi-spin systems and the implemented fitting routine}\label{Appendix:multi_spin_systems}
For simplicity, the main text considers only the case of one system spin.
The presented algorithm can however be generalized to cover also multi-spin systems.
An example of a multi-spin Hamiltonian is
\begin{align}
\hat H_\textrm{S} &= -\sum_{i \in \textrm{system}}\frac{\hbar\Delta_{ii}}{2}\hat\sigma^i_z \nonumber\\
&+ \frac{1}{2}\sum_{i<j \in \textrm{system}}\hbar\left(\Delta_{ij}\hat\sigma_+^i\hat\sigma_-^j + \Delta_{ij}^*\hat\sigma_+^j \hat\sigma_-^i \right) \, , \\
\hat H_\textrm{B} &= \sum_{k} \hbar\omega_{k} \hat b^\dagger_{k} \hat b_{k} \, , \\
\hat H_\textrm{C} &= \frac{1}{2}\sum_{i \in \textrm{system}} \hat\sigma_z^i\sum_{k} \left(v_{ik}^*\hat b^\dagger_k + v_{ik}\hat b_k\right)  \, .
\end{align}
The couplings $\Delta_{i\neq j}$ and $v$ may be complex numbers (as only a select number of phases can be absorbed into definitions of the boson operators).
This Hamiltonian is commonly used to describe exciton transport in photosynthesis,
where the system spins correspond to excitons and the bosonic modes to molecular vibrations.

The spectral function is defined analogously as for the single-spin system.
The multi-dimensional spectral function is defined as
\begin{align}\label{eq:multidimensional_spectral_function}
S_{ij}(\omega) = \int dt e^{i\omega t}\langle \hat X_i(t) \hat X_j(0)\rangle_0  \, ,
\end{align}
where
\begin{align}
X_i = \sum_{k}(v_{ik}\hat b_k + v_{ik}^*b^\dagger_k) \, .
\end{align}
In thermal equilibrium,
\begin{align}
S_{ij}(\omega) &= 2\pi\frac{ \sum_{k} v_{ik}v^*_{jk} \delta(\omega - \omega_k)}{1-\exp\left( -\frac{\hbar\omega}{k_{\rm B}T } \right)} \, ,\textrm{ for } \omega>0 \\
S_{ij}(\omega) &= 2\pi\frac{ \sum_{k} v^*_{ik}v_{jk} \delta(\omega + \omega_k)}{\exp\left( \frac{\hbar\omega}{k_{\rm B}T} \right) -1} \, ,\textrm{ for } \omega<0 \, , \\
S_{ji}(\omega) &= S_{ij}^*(\omega) \, .
\end{align}
We see that cross-correlations ($i\neq j$) can appear if different spins couple to the same bath modes.

The coarse graining is performed similarly for a single-spin system and a multi-spin system.
Here, the original bosonic bath is replaced by $n$ auxiliary boson modes.
Each of these auxiliary modes has a central frequency $\omega_m$,
coupling to the $i^\text{th}$ system spin $v_{im}$, and broadening $\kappa_m$.
For a single-spin system, the coarse-grained spectral function has the form~(\ref{eq:coarse_grained_spectral_function}).
For a multi-spin system this is generalized to
\begin{align}
S_{ij}(\omega) = \sum_{m=1}^{n} v_{im} v^*_{jm}\frac{\kappa_m}{(\kappa_m/2)^2+(\omega-\omega_m)^2} \, .
\end{align}

In the fitting presented in this paper, we optimize the cost function
\begin{align}
C= \sum_{i\leq j} \int_{\omega_\textrm{min}}^{\omega_\textrm{max}} d\omega \left\vert S_{ij}(\omega) - S^\textrm{target}_{ij}(\omega)\right\vert^2  \, ,
\end{align}
by seeking optimal values for $\omega_m$, $v_{im}$, and $\kappa_m$.
The fitting problem includes $n$ frequencies, $n n_\textrm{s}$ couplings, and $n$ broadenings,
where $n_\textrm{s}$ is the number of system spins.
Furthermore,
the system-qubit noise can be included similarly as in the case of the single-spin system (Sec.~\ref{sec:coarse_graining_scheme2}).
Here, a constant background rate is added to diagonal components of~$S_{ij}(\omega)$,
assuming the system-qubit noise is uncorrelated.
Possible differences in the system-qubit decoherence rates can be accounted for by
generalizing the constant~$r$ in Eq.~(\ref{eq:noise_ratio})
to a vector~$r_j$, where the index $j$~refers to a system spin.

The fitting is restricted to some relevant frequency window, here from $\omega_\textrm{min}$ to $\omega_\textrm{max}$,
which is usually determined by the energy levels of the system, the size of the system-bath couplings, and/or the frequency cut-off.
One should also keep in mind that the target spectral function, possibly obtained from an experiment,
may contain significant uncertainty, thereby limiting the attainable fitting accuracy.

The computational task of finding optimal parameters faces the general difficulties posed by non-linear fitting problems.
The optimization as presented above is a relatively easy task, but requires one to have a good heuristic for a starting guess.
If fitting in a black-box fashion, spectral functions that are sums of sharp peaks or are very broad represent simple problems.
But spectra with large gaps, or with a combination of broad and sharp peaks, are more challenging.
In this case, a good initial guess is crucial.
As the dimensionality grows, the parameter space grows exponentially.
However, for the purposes of this work, i.e., for noisy qubits of NISQ computers, one is in practice dealing with at most several tens of fitting modes,
and in this case suitable heuristics work most of the time, and the fitting fails only in pathological cases.

\section{Noisy-algorithm model}\label{Appendix:effective_noise_model} 
Here we give a short description of our model of quantum computing with incoherent error.
We assume that the noise inserted after each gate is qubit damping and dephasing.
This can be justified for hardware with such intrinsic noise and for the application of small-angle gates.
The form of the noise in the case of large-angle gates may vary~\cite{Fratus2023}.
However, the principles of deriving the effective noise model and optimizing it remain the same.
For a more detailed derivation and validity analysis of this noise model see Ref.~\cite{Fratus2022}.


\subsection{Noise after gates (physical noise)}\label{sec:model_of_qc}

For the following discussion, we assume that the time-propagation algorithm is based on a Trotter expansion over time $t=m\tau$,
\begin{align}
e^{-\textrm{i} \hat H t} = e^{-\textrm{i} \hat H m \tau} = \left[ e^{-\textrm{i} \hat H \tau} \right]^m \approx \left[ \Pi_{j} e^{-\textrm{i} \hat H_j \tau} \right]^m  \, .
\end{align}
Here the Hamiltonian $H$ is divided into partial Hamiltonians,
\begin{align}
    \hat H=\sum_j \hat H_j \, ,
\end{align}
whose contributions to the time evolution are implemented using available unitary gates.
For native (or natural) gates we have
\begin{align}
    e^{-\textrm{i} \hat H_j \tau} = \hat U_j \, ,
\end{align}
whereas non-native decompositions have the form
\begin{align}\label{eq:decomposition_block}
    e^{-\textrm{i} \hat H_j \tau} = \Pi_k \hat U_{j, k}  \, .
\end{align}
In practice, the elementary gates $\hat U_{j,k}$ include large-angle operations,
such as $R_x(\pi/2)$ or CNOT.

In our modeling,
the physical noise is included as non-unitary operations after unitary gates,
\begin{align}
    \hat U  &\rightarrow  {\cal N} {\cal U} \, .
\end{align}
On the right-hand side, the unitary gate is represented as a superoperator ${\cal U}$
and the non-unitary noise as a Kraus operator ${\cal N}$.
It is always possible to establish such description of an incoherent error,
assuming it keeps qubits in their computational basis.
In our modeling, also the identity gate is assumed to come with noise.
In other words, the noise is acting also on idling qubits.

Assuming the effect of the noise per gate is weak,
it is possible replace the Kraus superoperators by Lindbladians~${\cal L}_j$
and physical gate times~$t_j$ such that
\begin{align}
    {\cal N}_j &\approx 1 + t_j {\cal L}_j \, ,
\end{align}
where ${\cal L}_j$ is a Lindblad operator describing the physical noise.

\subsection{Effective noise}\label{sec:noise_mapping}

Noise after the unitary gates can appear in a different form in the simulated system~\cite{Fratus2022}.
To understand how this noise mapping works,
consider a simple example of decomposing some unitary to three elementary gates, $\hat V= \hat U_3 \hat U_2 \hat U_1$.
Since unitary gates have also inverse gates, we can rewrite the noisy version of this circuit as
\begin{align}
    \hat U_3 \hat U_2 \hat U_1 &\rightarrow{\cal N}_3 {\cal U}_3 {\cal N}_2 {\cal U}_2 {\cal N}_1 {\cal U}_1  \nonumber \\
    &= {\cal N}_3 {\cal U}_3 {\cal N}_2 {\cal U}_3^{-1} {\cal U}_3 {\cal U}_2 {\cal N}_1 {\cal U}_2^{-1} {\cal U}_3^{-1} {\cal U}_3 {\cal U}_2 {\cal U}_1 \nonumber \\
    & \equiv {\cal N}_3 {\cal N}_2' {\cal N}_1' {\cal U}_3 {\cal U}_2 {\cal U}_1 \nonumber \\
    &\equiv {\cal N} {\cal V} \, ,
\end{align}
where ${\cal N} =  {\cal N}_3 {\cal N}_2' {\cal N}_1'$ and
\begin{align}
    {\cal N}'_1 &= {\cal U}_3 {\cal U}_2 {\cal N}_1 {\cal U}_2^{-1} {\cal U}_3^{-1} \\
    {\cal N}'_2 &= {\cal U}_3 {\cal N}_2 {\cal U}_3^{-1}  \, .
\end{align}
We see that the noise superoperator ${\cal N}_1$ got rotated by the unitary transformation ${\cal U}_3{\cal U}_2$
and the noise superoperator ${\cal N}_2$ by ${\cal U}_3$.
Furthermore, under the assumption of small noise-probability
we can approximate
\begin{align}
    {\cal N}_3 {\cal N}_2' {\cal N}_1' &\approx (1+t_{3}{\cal L}_3)(1+t_{2} {\cal L}'_2)(1+t_{1} {\cal L}'_1) \nonumber \\
    &\approx 1 + t_{1} {\cal L}'_1 + t_{2} {\cal L}'_2 + t_{3} {\cal L}_3  \, .
\end{align}
Here, the operators in the individual Lindbladians ${\cal L}_i$ are rotated by unitary gates $\hat U_{j>i}$
exactly in the same way as the noise superoperators were rotated by ${\cal U}_{j>i}$.
The (noise part of) the effective Lindbladian then becomes
\begin{align}
\tau {\cal L}_\textrm{eff} &= t_{1} {\cal L}'_1 + t_{2} {\cal L}'_2 + t_{3} {\cal L}_3 \, .
\end{align}
This result can be generalized to arbitrary circuits~\cite{Fratus2022}.

The transformation of noise to other effective forms in a series of unitary gates
has also been studied recently in Refs.~\cite{Wang2020, Sommer2021, Sun2021}.
The effect of quasistatic noise in digital quantum simulation has been addressed in Ref.~\cite{Reiner2018}.

\subsection{Noise transformations in the examples}

\begin{figure}
\includegraphics[width=0.8\columnwidth]{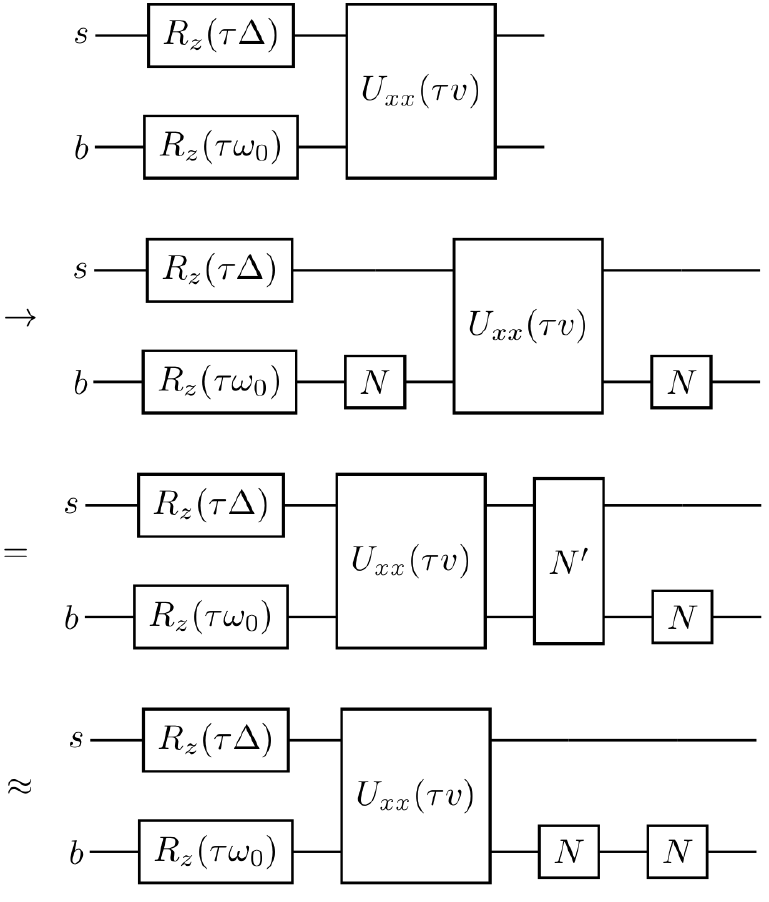}
\caption{Derivation of the effective noise model in the case of native gates and a
circuit that describes a system-spin $s$ coupled to a bath-spin $b$ via XX-interaction.
Only the bath qubit is noisy. In the derivation, the noise operators are commuted to the end of the circuit.
In the case of $\tau v \ll 1$ we have $\hat N'\approx \hat N$.
}\label{fig:circuit1}
\end{figure}
In most examples we consider native XX-interaction (variable MS) gates.
In this case, the effective noise and the physical noise have the same form,
since we have approximately
\begin{align}
\hat N' = \hat U_{xx}(\tau v) \hat N \hat U_{xx}(-\tau v) &\approx \hat N \, ,
\end{align}
which is true when $\tau v \ll 1$ and $\hat U_{xx}$ is a native gate.
The derivation of this result is sketched in Fig.~\ref{fig:circuit1}.

\begin{table*}
\begin{center}
    \begin{tabular}{c | c | c | c | c | c | c }
    \multicolumn{1}{c}{} & \multicolumn{1}{|c|}{$\sigma_-^\textrm{control}$}  & $\sigma_+^\textrm{control}$ & $\sigma_z^\textrm{control}$ &  $\sigma_-^\textrm{target}$ & $\sigma_+^\textrm{target}$ & $\sigma_z^\textrm{target}$  \\ [0.5ex]
    \hline \hline
    CNOT & $\sigma^\textrm{control}_-\sigma_x^\textrm{target}$  & $\sigma^\textrm{control}_+\sigma_x^\textrm{target}$  & $\sigma_z^\textrm{control}$ & $P_0 \sigma_-^\textrm{target} + P_1 \sigma_+^\textrm{target}$ &  $P_0 \sigma_+^\textrm{target} + P_1 \sigma_-^\textrm{target}$  & $\sigma_z^\textrm{control} \sigma_z^\textrm{target}$ \\ 
    \hline
    Control-Z & $\sigma_-^\textrm{control}\sigma_z^\textrm{target}$ & $\sigma_+^\textrm{control}\sigma_z^\textrm{target}$  & $\sigma_z^\textrm{control}$ &  $\sigma_-^\textrm{target}\sigma_z^\textrm{control}$ & $\sigma_+^\textrm{target}\sigma_z^\textrm{control}$ & $\sigma_z^\textrm{target}$ \\ 
    \hline
\end{tabular}
\end{center}
\caption{Effective noise for different incoming noise types and two-qubit gates. We consider physical noise operators $\hat\sigma_-,\hat\sigma_+,\hat\sigma_z$
under control-X (CNOT) and control-Z unitary transformations.
Here
$\hat P_0=(1+\hat\sigma^\textrm{control}_z)/2$ and $\hat P_1=(1 - \hat\sigma^\textrm{control}_z)/2$
are control-qubit projection operators.
}\label{table:rotations_of_operators}
\end{table*}

The derivation of the effective noise for non-native decompositions is done as follows.
In the case of CNOT-B decomposition of the XX-interaction (with bath qubits as the control qubits), see Sec.~\ref{sec:spin_coupled_to_one_mode},
we look how the control-qubit~$i$ damping and dephasing operators transform under CNOT, see Table~\ref{table:rotations_of_operators},
\begin{align}
\textrm{CNOT}\ \hat\sigma_-^i\ \textrm{CNOT} &= \hat\sigma_-^{i} \hat\sigma_x  \\
\textrm{CNOT}\ \hat\sigma_z^i\ \textrm{CNOT} &= \hat\sigma_z^{i} 
\end{align}
Remember that a superscript~$i$ in a Pauli operator refers to a bath qubit (or the corresponding auxiliary spin).
This tells us that physical bath-qubit damping gets translated into simultaneous bath-spin damping and system-spin flip, whereas dephasing keeps its form.
This result means that the effective Lindbladian has the form
\begin{align}\label{eq:LindbladianCNOT1}
{\cal L}_\textrm{eff}[\hat \rho] &\approx \frac{\textrm{i}}{\hbar}[\hat\rho, \hat H] + \left(\gamma-\delta\gamma\right)\sum_{i}{\cal L}_{\sigma_-^i}[\hat \rho] + \frac{\Gamma}{2}\sum_{i}{\cal L}_{\sigma_z^i}[\rho] \nonumber \\
&+ \delta\gamma\sum_{i}{\cal L}_{\sigma_-^i\sigma_x}[\hat \rho] \, ,
\end{align}
where we use a notation ${\cal L}_{A}$ for a noise Lindbladian with noise operator $\hat A$.
It turns out that the last term on the right-hand side of Eq.~(\ref{eq:LindbladianCNOT1})
can be approximately factored into two uncorrelated contributions,
\begin{align}
{\cal L}_\textrm{eff}[\hat \rho] &\approx \frac{\textrm{i}}{\hbar}[\hat\rho, \hat H] + \left(\gamma-\delta\gamma\right)\sum_{i}{\cal L}_{\sigma_-^i}[\hat \rho] + \frac{\Gamma}{2}\sum_{i}{\cal L}_{\sigma_z^i}[\hat \rho] \nonumber \\
 &+ \delta\gamma\sum_{i}{\cal L}_{\sigma_-^i}[\hat \rho] + \gamma_\textrm{system} {\cal L}_{\sigma_x}[\hat \rho] \, ,
\end{align}
where
\begin{align}
\gamma_\textrm{system} &= \delta\gamma\sum_{i}\langle \hat\sigma_+^i \hat\sigma_-^i \rangle \, .
\end{align}
We also get $\gamma_\textrm{system} \ll \gamma, \Gamma$,
following from the fact that most of the time the individual bath qubits are at ground. 
It follows that $\gamma_\textrm{system}$ can be neglected and we have
\begin{align}
{\cal L}_\textrm{eff}[\hat \rho] &\approx \frac{\textrm{i}}{\hbar}[\hat\rho, \hat H] + \gamma\sum_{i}{\cal L}_{\sigma_-^i}[\hat \rho] + \frac{\Gamma}{2}\sum_{i}{\cal L}_{\sigma_z^i}[\hat \rho] \, ,
\end{align}
This has the same form as the physical noise.

For CNOT-S, {\it i.e.}, when the system qubit is the control qubit, the changes in the effective model are larger.
This can be derived by looking at the target-qubit noise conversions, see Table~\ref{table:rotations_of_operators},
\begin{align}
&\textrm{CNOT}\ \hat\sigma_-^{i}\ \textrm{CNOT} = \hat P_0 \hat\sigma_-^{i} + \hat P_1 \hat\sigma_+^{i} \\
&\textrm{CNOT}\ \hat\sigma_z^{i}\ \textrm{CNOT} = \hat\sigma_z \hat\sigma_z^{i} \, ,
\end{align}
where  $\hat P_0=(1+\hat\sigma_z)/2$ and $\hat P_1=(1 - \hat\sigma_z)/2$ are projection operators of the system qubit.
The first equation implies that physical bath-qubit decay can also transform into effective bath-spin excitation.
It can therefore be active even when the bath is empty.
Furthermore, this correlated noise wants to project the system spin to one of its $\hat\sigma_z$ eigenstates.

The effective Lindbladian becomes
\begin{align}
{\cal L}_\textrm{eff}[\hat \rho] & \approx \frac{\textrm{i}}{\hbar}[\hat\rho, \hat H] + \gamma\sum_{i}{\cal L}_{\sigma_-^i}[\hat \rho] + \frac{\Gamma}{2}\sum_{i}{\cal L}_{\sigma^z_i}[\hat \rho] \nonumber \\
& +\frac{\bar \Gamma}{2} {\cal L}_{\sigma_z}[\hat \rho] \, .
\end{align}
where we have done noise factoring and neglected a contribution introducing bath excitation (this approximation was verified numerically).
The size of $\bar \Gamma$ can be solved numerically and gives a noticeable system-spin dephasing.

The effective noise analysis of the optimal control-Z decomposition (Fig.~\ref{fig:circuit3}) is essentially the same as made for the CNOT-B decomposition
(see also Table~\ref{table:rotations_of_operators}).
This means that the effective noise of a control-Z decomposition has the same form as the physical noise.


\section{Quantum circuit optimization}\label{Appendix:optimized_algorithms}
We assume that the effective noise given by~${\cal L}_\textrm{eff}$ is dominated by spin damping and spin dephasing.
As discussed above, this is possible for circuit decompositions based on native (small-angle) decompositions.
However, if large-angle gates are needed, the quantum circuits should be designed so that
the large angle gates act only on noiseless system qubits, or one uses large-angle gates that do not modify the noise.
Also, the simulation algorithm used (like the swap-network algorithm) should ideally support this approach.
Furthermore, if also the system qubits are noisy, it may be beneficial to transform the system noise to a symmetric form,
in order to map it to an open-system model. These aspects are discussed in detail in this Appendix.

\subsection{Avoiding large-angle bath-qubit rotations}

Large-angle rotations of bath qubits may rotate the noise operators $\sigma_+^i$ and $\sigma_z^i$ to very different forms.
Such rotations should then clearly be avoided.
We give below several examples of gate decompositions that follow this principle.

\begin{figure}
\includegraphics[width=0.93\columnwidth]{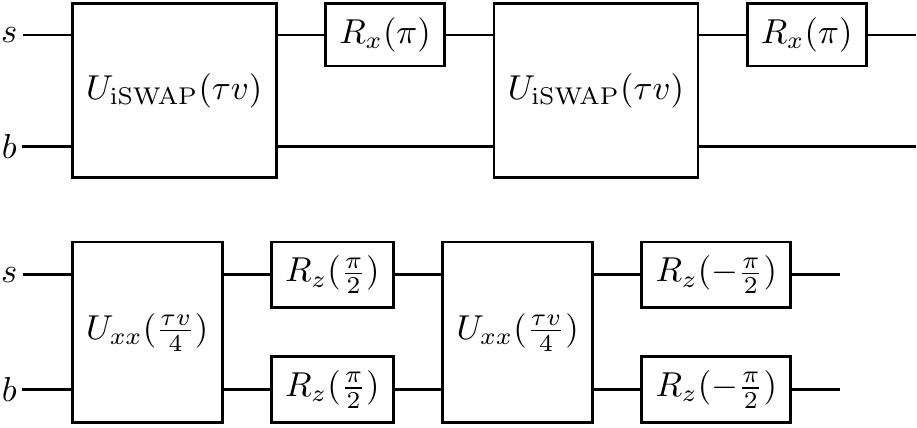}
\caption{Examples of decomposing spin-spin interactions by native gates, without causing alternations to the effective noise model.
(Above circuit) XX-interaction decomposed by variable iSWAP gates. (Below circuit) Excitation hop decomposed by variable XX-gates.
}\label{fig:circuit2}
\end{figure}

Ideally we have access to a native gate of some two-qubit interaction.
In this paper we construct time-propagation according to XX-interaction between a system qubit and a bath qubit.
Consider realizing this using a variable iSWAP two-qubit gate, which is a native gate of an excitation hop,
\begin{align} 
\hat U_{\textrm{iSWAP}}(\tau v) = e^{-\textrm{i}(\hat\sigma_+^s\hat\sigma_-^b + \hat\sigma_-^s\hat\sigma_+^b) \tau v/2} \, .
\end{align}
Here 
$s$ refers to a system qubit and $b$ to a bath qubit.
To obtain the full XX-interaction, we need to add the counter-rotating terms.
We can create these terms by another variable iSWAP gate by surrounding it by X-gates of one qubit.
To avoid rotations of the bath qubit, the X-gates are applied to the system qubit,
\begin{align}
\hat R_{x,s}(\pi) \hat U_{\textrm{iSWAP}} (\tau v) \hat R_{x,s}(\pi) &= e^{-\textrm{i}(\hat\sigma_+^s\hat\sigma_+^b + \hat\sigma_-^s\hat\sigma_-^b) \tau v/2} \, .
\end{align}
The time propagation is then generated by the decomposition shown in Fig.~\ref{fig:circuit2} (above circuit).
When using this construction, and assuming noiseless system qubits,
the physical and effective noise Lindbladians have the same form.
Note that
this is not the case for the decomposition where X-gates are performed on the bath qubits.

On the other hand,
to obtain an excitation hop using the native XX-gate 
\begin{align}
\hat U_{xx}(\tau v) & = e^{-\textrm{i} \hat\sigma^s_x\hat\sigma^b_x \tau v/2} \, ,
\end{align}
we need to supplement this with YY interaction.
This can be created from an XX-gate with the help of Z-rotations,
\begin{align}
& \hat R_{z,s}\left(-\frac{\pi}{2}\right) \hat R_{z,b}\left(-\frac{\pi}{2}\right) U_{xx}(\tau v) R_{z,b} \left(\frac{\pi}{2}\right) R_{z,s} \left(\frac{\pi}{2}\right) \nonumber \\
&= e^{-\textrm{i} \hat\sigma^s_y\hat\sigma^b_y \tau v/2}  \, .
\end{align}
Note that this also includes large-angle rotations of the bath qubit.
However, Z-rotations do not change the form of damping ($\hat\sigma_-$) or dephasing ($\hat\sigma^z$) operators
(they just introduce a phase shift which cancels out in the Lindbladian).
The excitation hop is then generated by the decomposition shown in Fig.~\ref{fig:circuit2} (below circuit).

\begin{figure}
\includegraphics[width=0.9\columnwidth]{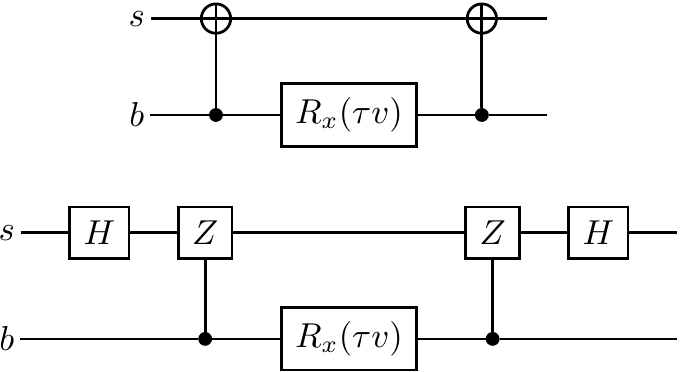}
\caption{Optimal CNOT (above) and control-Z (below) decomposition of the XX-interaction. Large-angle rotations are performed only to system qubits,
which is expected to reduce the alternation of the effective noise.
}\label{fig:circuit3}
\end{figure}

Also non-native two-qubit gate decompositions should follow this principle.
For example, according to our results in Sec.~\ref{sec:spin_coupled_to_one_mode}, using the bath qubit as a CNOT control-qubit seems beneficial.
This can be understood qualitatively as a result of the fact that the bath qubit is never flipped (subjected to a large-angle gate).
This decomposition is shown in Fig.~\ref{fig:circuit3} (above circuit).
In Fig.~\ref{fig:circuit3} is also shown the optimal decomposition when the control-Z gate is used~(below circuit).

\subsection{System-bath swap-network}

In common open-system models the bath is non-interacting.
For the presented quantum algorithm
this means that a direct connectivity between the bath qubits is not needed.
Rather a connectivity from the system to all bath qubits.
If this is provided,
a simple Trotter expansion of the time-propagation is adequate.

\begin{figure}
\includegraphics[width=\columnwidth]{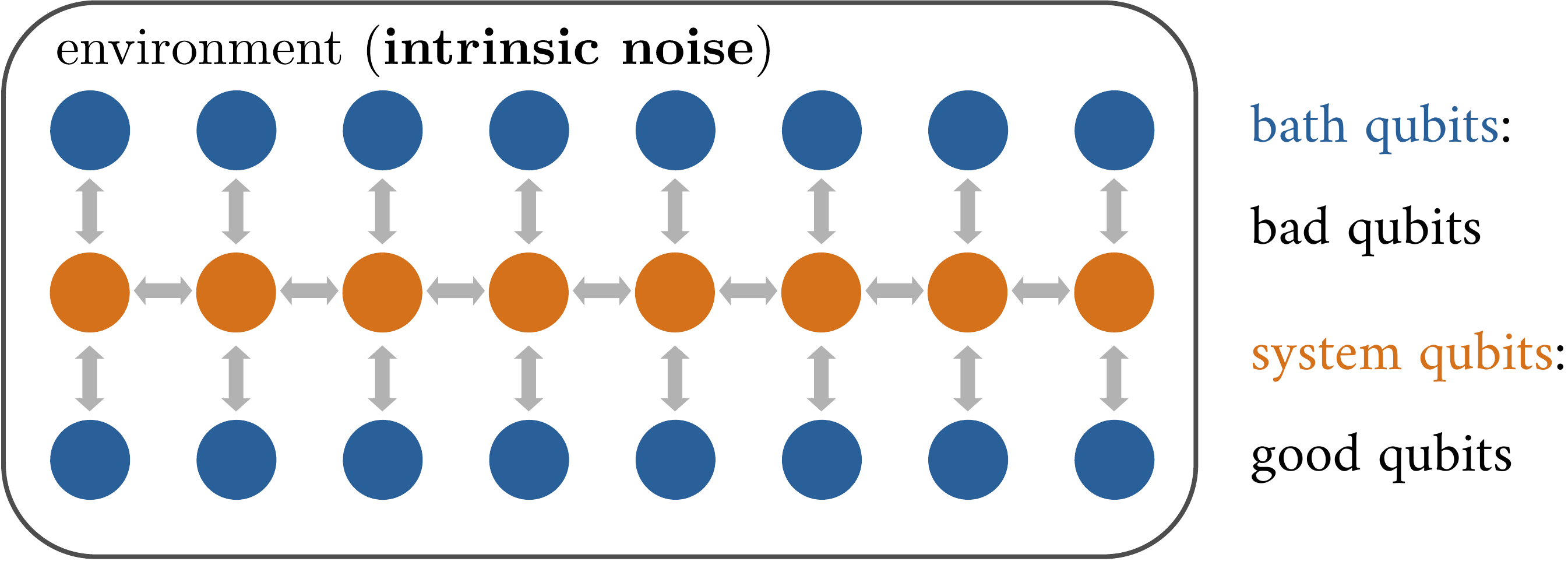}
\caption{A two-dimensional qubit architecture that is optimal for performing a Trotter time evolution of a system-bath model.
In total $n_\textrm{s}$~system qubits locate between $2n_\textrm{s}$ bath qubits.
The bath qubits can also refer to any other resonance modes in the device,
whose interaction with the system qubits can be controlled digitally.
}\label{fig:figure0}
\end{figure}

\begin{figure}
\includegraphics[width=0.6\columnwidth]{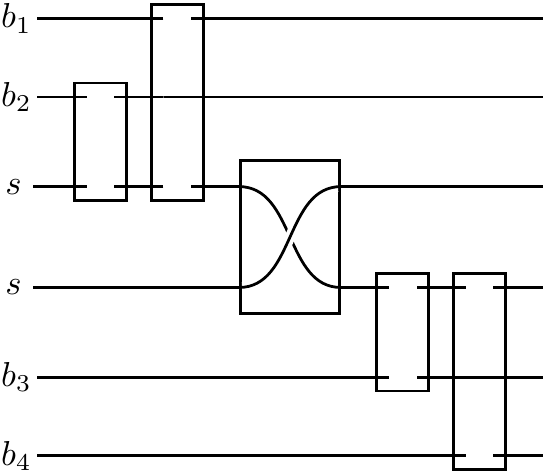}
\caption{Example of a system-bath swap algorithm on a two-dimensional qubit architecture.
Two system qubits are used to store the information of one system spin.
The system spin interacts with four auxiliary spins, represented by the four bath qubits.}
\label{fig:circuit4}
\end{figure}

On the other hand, if only nearest-neighbor interactions are possible, but the device has a two-dimensional architecture,
an efficient swap-network algorithm can be developed.
Here the quantum states of the system spins are moved (swapped) in the system-qubit network,
located between the bath qubits, see Fig.~\ref{fig:figure0}.
The difference to the common swap algorithm (see for example~\cite{Kivlichan2018}) is that bath states are not swapped.
Within this modification, one avoids doing large-angle rotations of the bath qubits.
In the example circuit shown in Fig.~\ref{fig:circuit4}, the state of one system-spin is stored by two system qubits, one per time.
The system spin interacts with four auxiliary spins, represented by the four bath qubits.
Note that this method most optimally time-propagates $n_\textrm{s}$ system spins coupled to $2n_\textrm{s}$ auxiliary spins.

\subsection{Noise symmetrization}\label{sec:optimal_algorithms}
So far we have assumed that the system qubits are noiseless.
Finite system noise may also be included, if it is transformed into the proper form.
This can be achieved by noise symmetrization, whose goal is to map system noise to heating of the modeled bath.

\begin{figure}
\includegraphics[width=0.8\columnwidth]{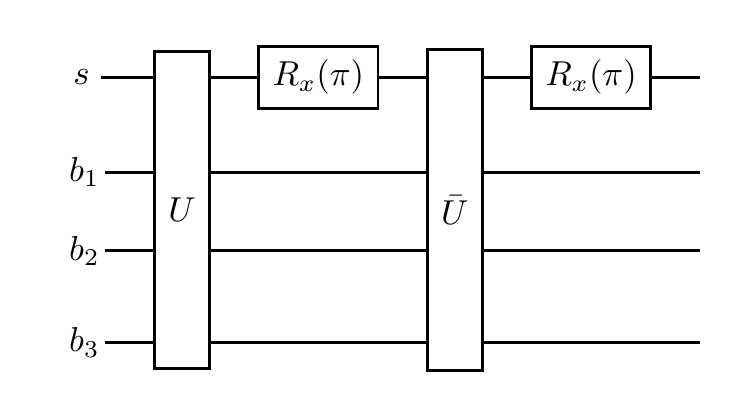}
\caption{Example of a noise-symmetrization algorithm based on application of X-gates to the system qubit between the Trotter steps.
Here $\hat {\bar U} = \hat R_{x,s}(\pi) \hat U \hat R_{x,s}(\pi)$.
The goal is the transformation of system-qubit damping partially into effective excitation,
which allows it to be mapped on the bath spectral function.
}
\label{fig:circuit5}
\end{figure}

A simple example of noise symmetrization is the application of
X-gates to transform physical qubit-damping into effective spin-excitation.
This is based on the transformation
\begin{align}
\hat\sigma_x \hat\sigma_-\hat\sigma_x = \hat\sigma_+ \, .
\end{align}
We then insert X-gates between (original) Trotter-steps,
and make corresponding changes to the unitary gates (such as $Y\rightarrow -Y$).
In the case of system-qubit damping symmetrization,
we then have the new Trotter-step circuits as shown in Fig.~\ref{fig:circuit5}.
Here $\hat {\bar U} = \hat R_{x,s}(\pi) \hat U \hat R_{x,s}(\pi)$.
Ideally the decomposition of $\hat {\bar U}$ is similar to that of $\hat U$, or even $\hat {\bar U} = \bar U$, which would be the case for an XX-interaction.
This trick symmetrizes the effective Lindbladian~${\cal L}_\textrm{eff}$ terms as
\begin{align}
\gamma\hat\sigma_-\hat \rho\hat\sigma_+ &\rightarrow \frac{\gamma}{2}\hat\sigma_-\hat \rho\hat\sigma_+ + \frac{\gamma}{2}\hat\sigma_+\hat \rho\hat\sigma_- \, .
\end{align}
Furthermore, since
\begin{align}\label{eq:symmetrization}
\frac{\gamma}{2}\hat\sigma_-\hat \rho\hat\sigma_+ + \frac{\gamma}{2}\hat\sigma_+\hat \rho\hat\sigma_- &=\frac{\gamma}{4}\hat\sigma_X \hat \rho \hat\sigma_X + \frac{\gamma}{4}\hat\sigma_Y \hat \rho \hat\sigma_Y \, ,
\end{align}
we see that damping has changed to a coupling to two independent baths, via operators $\hat\sigma_x$ and $\hat\sigma_y$.
As demonstrated in Sec.~\ref{sec:SET} and discussed more in Appendix~\ref{Appendix:bosons_from_fermions},
this can be mapped to a constant background of spectral functions in a certain interesting class of open system models.

The system-noise symmetrization can be generalized to all spin-directions,
in which case the system-noise becomes equivalent to depolarization.
This corresponds to coupling to three independent baths,
via operators $\hat\sigma_x$, $\hat\sigma_y$, and $\hat\sigma_z$.

It should be noted that damping during the X-gate itself may transform partly into $\hat\sigma_z$-noise
in terms of the circuit-level noise that is inserted after the X-gates~\cite{Fratus2023},
in which case the bath mapping to $\hat\sigma_x$ and $\hat\sigma_y$ noise (shown above) is only approximate.
However, in the case of long circuit depths, the noise contribution from the X-gates themselves is small.
Various noise symmetrization approaches have been studied experimentally in Refs.~\cite{Rost2020, Sun2021, Hashim2021}.

\section{Error sources}\label{sec:error_and_optimization_analysis}

Here we discuss three main error sources in the quantum algorithm.
These are the Trotter error, the coarse-graining error, and the bath-Gaussianity error.
We also discuss how the key parameters determining their sizes become connected in the quantum algorithm.

\subsection{Trotter error}
The time propagation is realized in time steps~$\tau$.
Unless all terms in the Trotter expansion commute, there is an error in the time propagation~\cite{Childs2021}. 
At the center of this analysis is the BCH (Baker-Campbell-Hausdorff) formula, which states
\begin{align}\label{eq:BCH}
\Pi_i \exp\left(\tau\hat A_i\right) = \exp\left(\tau \hat Z\right) \, ,
\end{align}
where to first order in $\tau$ we have
\begin{align}
\tau\hat Z &= \sum_i \tau\hat A_i + \frac{\tau^2}{2}\sum_{i<j}[\hat A_i, \hat A_j] \, .
\end{align}
The operator $\frac{\tau^2}{2}\sum_{i<j}[\hat A_i, \hat A_j]$ represents an error in the Trotterization.
If Eq.~(\ref{eq:BCH}) corresponds to one Trotter step,
the total accumulated error during the full time evolution can now be estimated to be
\begin{align}
\epsilon_\textrm{Trotter} \sim {\cal O}\left( m \tau^2\left\vert\left\vert \sum_{i<j}[\hat A_i, \hat A_j]  \right\vert\right\vert \right) \, ,
\end{align}
where $m$ is the number of Trotter steps and the total simulated time $t=m\tau$.
This is of course a rather rough estimate and should be analyzed more carefully for specific cases.
Without the presence of noise, the operators $\hat A_i$ can be identified as exponentiated partial Hamiltonians.
When including the noise, it is more convenient to switch to the superoperator notation,
where the operators $\hat A_i$ are exponentiated commutators (the gates) and exponentiated Lindbladians (the noise)~\cite{Fratus2023}.

Let us first look closer at pure gate contributions to the Trotter error.
Assuming we are using the first-order Trotterization formula,
we find that the error can be interpreted as an additional Hamiltonian term $\tilde H$ in the effectively time-propagated model (considering now single-spin system, single-mode bath):
\begin{align}\label{eq:example2}
& \exp\left[ - \frac{\textrm{i} \tau}{\hbar}\left( \hat H_\textrm{S} + \hat H_\textrm{B}\right) \right] \exp\left( - \frac{\textrm{i} \tau}{\hbar} \hat H_\textrm{C}\right) \nonumber \\
&= \exp\left[ - \frac{\textrm{i} \tau}{\hbar}\left( \hat H + \tilde H\right) \right] \\
\tilde H &= v\tau\left( -\frac{\Delta}{4} \hat \sigma_y \sum_{j=1}^N \frac{\hat\sigma_x^{j}}{\sqrt{N}} - \frac{\omega_0}{4} \hat \sigma_x \sum_{j=1}^N \frac{\hat\sigma_y^{j}}{\sqrt{N}}\right)  \, .
\end{align}
When comparing the form of $\tilde H$ to the original Hamiltonian~$\hat H$,
we can interpret each coupling term as possessing an error term with characteristic size
$\sim {\cal O}\left[v\tau\left(\vert\Delta\vert + \omega_0\right)\right]$.
We want this contribution to be small.
We can compare this to the energy scale of the original coupling term.
Assuming that $\Delta\sim\omega_0$, the error has a relative magnitude ${\cal O}(\Delta\tau)$.
We would then demand
$\Delta\tau  \ll 1$.
Similarly, if we compare the error to the scale of the system and bath energies, we get demand $v\tau  \ll 1$.
In general a comparison to the energy scale of the (possibly slow) solution dynamics may need to be made,
which can differ from the energy scale of the Hamiltonian.
In our case, the dissipation (with rate $\kappa$) however sets a limit for long-time correlations.

Special to our algorithm is the Trotter error in the noise terms. This follows the same math as above.
It is however now more convenient to work with superoperators.
In this paper, qubit noise is assumed to be uncorrelated and thereby
the relevant contribution comes from commuting noise with gates.
We consider now only commuting bath noise with gates that reproduce the $\hat \sigma_x\hat \sigma_x^i$-interaction (where $i\in\textrm{bath}$),
since the noise (damping or dephasing) superoperator commutes with the gates producing the $\hat \sigma_z^i$ terms.
A commutator between noise and gate terms generates another noise superoperator,
whose ``normal'' operators are commuted with corresponding partial (gate) Hamiltonians:
\begin{align}\label{eq:Trotter_error_Lindblad}
\exp\left( {\cal L}_\textrm{eff}\tau \right) & \rightarrow \exp\left[ \left( {\cal L_\textrm{eff}} + \sum_i\tilde {\cal L}^i_\textrm{eff} \right)\tau \right] \\
\tilde {\cal L}^i_\textrm{eff}[\rho] &= \left[\hat O_i, \hat\sigma_-^i \right]\hat \rho \hat\sigma_+^i + \hat\sigma_-^i\hat \rho \left[\hat O_i, \hat\sigma_+^i \right] \nonumber \\
&-\frac{1}{2}\left(\left[\hat O_i, \hat\sigma_+^i\hat\sigma_-^i\right]\hat\rho + \rho\left[\hat O_i, \hat\sigma_+^i\hat\sigma_-^i\right] \right) \\
\hat O_i &= \gamma\tau\frac{-\textrm{i}v}{4\hbar}  \hat \sigma_x\hat \sigma_x^i \, .
\end{align}
Here we explicitly assume damping noise.
We find $\left[\hat O_i, \hat\sigma_+^i \right] \propto \hat \sigma_x\hat \sigma_z^i$,
 $\left[\hat O_i, \hat\sigma_-^i \right] \propto -\hat \sigma_x\hat \sigma_z^i$, and
$\left[\hat O_i, \hat\sigma_+^i\hat\sigma_-^i \right] \propto \textrm{i}\hat \sigma_x\hat \sigma_y^i$.
The error term $\sum_i\tilde {\cal L}^i_\textrm{eff}$ then describes correlated effective noise.
Note that it has the same form as the leading-order term in $\tau$ when the Lindblad noise operators $\sqrt{\gamma}\hat \sigma_-^i$ are
rotated by $\exp(-\textrm{i}\tau \hat H_i/2\hbar)$.
The error has characteristic size
\begin{align}\label{eq:example4}
\epsilon_{\cal L} &\sim {\cal O}(v\tau\gamma) \, .
\end{align}
To have a small contribution in comparison to the correct noise Lindbladian, which has the magnitude~$\gamma$,
we then obtain the requirement~$v\tau \ll 1$.

The true size of the Trotter error in comparison to the relevant energy scales
is however hard to predict analytically and in practice one needs to resort to numerical simulations.
In the examples of Sec.~\ref{sec:examples}, we found that values~$v\tau\sim 0.2$ were small enough to have a negligible Trotter error on the visual scale of the plots.





\subsection{Coarse-graining error}
Unless the target spectral function is Lorentzian, or a (finite) sum of Lorentzians, there is always an error in the coarse graining.
Obviously, such an error is reduced when the number of lorentzians~$n$ is increased.
Several recent works study errors originating in the imperfect spectral-function fitting, as well as their correction,
in the context of classical numerical methods~\cite{Dorda2014, Mascherpa2017, Mascherpa2020}.
Particularly interesting for the noise-utilizing quantum algorithm is
an error-correction approach based on calculating functional derivatives with respect to spectral-function changes~\cite{Dorda2014}.
For the quantum algorithm considered here,
this would correspond to using a low-noise qubit as a narrow-peak perturbation in the spectral function,
to implement a functional derivative.
The leading-order correction is the derivative multiplied by the error in the spectrum~\cite{Dorda2014}.

\subsection{Bath-Gaussianity error} 
The bath Gaussianity error emerges when the auxiliary spins do not perfectly represent a bosonic bath.
In other words, the mapping of Eqs.~(\ref{eq:spin_boson_operator_mapping1}-\ref{eq:spin_boson_operator_mapping2}) leads to an error.
The error is generally of order ${\cal O}(1/N)$, where $N$ is the number of spins representing one bosonic mode.
For example, the operator $\hat s = \sum_i \hat\sigma_+^i \sum_i \hat\sigma_-^i/N$ has expectation values
$s + s(s-1)/N$ (in the relevant space of Dicke states~\cite{Henschel2010}), where $s$ is a non-negative integer.
The difference to the corresponding bosonic result is the term~$s(s - 1)/N$, which vanishes in the limit $N\rightarrow \infty$.
One should note that the actual size of this error can be studied in the quantum simulation by additional measurements and,
if needed, can be reduced by increasing the number of auxiliary spins used to represent one bosonic mode.

\subsection{Error-parameter dependency}
There is a special dependency between the central parameters defining the size of the different errors.
We derive now a connection for a simple example.
We start by noting that the coarse graining is always done within some frequency window, for example, within a cutoff frequency~$\omega_\textrm{c}$.
In the simplest (yet still reasonably accurate) estimate,
the auxiliary boson modes are inserted at constant frequency intervals with broadenings
\begin{align}\label{eq:tradeoff1}
\kappa \approx \frac{\omega_\textrm{c}}{n} \, .
\end{align}
We already note that an increase of $n$ decreases $\kappa$.
The circuit depth also depends on $n$ and for the models considered in this paper it is approximately
\begin{align}
D &\approx n N D_0 = n_\textrm{q} D_0 \label{eq:tradeoff2} \, ,
\end{align}
where $N$ is the number of qubits representing an auxiliary boson mode and for simplicity
we assume that it is the same for every boson mode.
The constant $D_0$ is defined by the simulation algorithm and the gate decomposition
(see Table~\ref{table:circuit_depths}). 
Combining the above formulas with the earlier result for the correct Trotter time step in the quantum simulation (Sec.~\ref{sec:noise_model_matching}),
\begin{align}
\tau &= \epsilon \frac{D}{\kappa} \, , \nonumber
\end{align}
leads to a special relation
\begin{align}\label{eq:error_tradeoff}
\tau \omega_\textrm{c}    &\approx n^2 N \epsilon D_0 \, .
\end{align}
The left-hand side of this equation is a measure of the Trotter error (on the energy scale of the Hamiltonian).
On the right-hand side, we have a multiplication between the square of the number of Lorentzians used in the coarse graining~$n$ (defining the coarse-graining error)
and the number of bath qubits per boson mode $N$ (defining the bath-Gaussianity error).
The factors $D_0$, $\omega_\textrm{c}$, and $\epsilon$ can be taken to be fixed constants.
A change in one parameter needs to be compensated with a change in the other parameters.
For example, achieving an increase in fitting accuracy by dropping $N$ from some previous value $N_0$ to~1, means a change $n^2 N =n^2N_0 \rightarrow (nN_0)^2 \times 1=n^2 N_0^2$,
leading to an additional factor of $N_0$ cost increase in the Trotter time step size.
We can also generally conclude that the lower the gate error~$\epsilon$, the better the quantum algorithm will perform,
since this allows for a larger tolerance in the other parameters ($n, N$, and $D_0$ can be larger, while $\tau$ can be smaller).

\section{Spin-boson model of electronic transport}\label{Appendix:bosons_from_fermions}

In this Appendix, we discuss how to approximately map a fermionic open-system model to a spin-boson model.
The key steps are (i) the assumption of Gaussian statistics of the fermion coupling operator,
(ii) the replacement of this operator by a boson coupling operator,
and (iii) the matching of the spectral functions.
It should be noted that this procedure is not restricted to the specific Hamiltonian considered below
but is also more generally applicable.

Our open quantum system is described by the Hamiltonian $\hat H = \hat H_\textrm{S} + \hat H_\textrm{B} + \hat H_\textrm{C}$, where
\begin{align}\label{eq:HamiltonianSET}
\hat H_\textrm{S} & = -\frac{\hbar\Delta}{2}\hat\sigma_z \, , \\
\hat H_\textrm{B} & = \sum_m\sum_{k} E_{km} \hat c^\dagger_{km} \hat c_{km} + \sum_m\sum_{l} E_{lm} \hat d^\dagger_{lm} \hat d_{lm} \, , \\
\hat H_\textrm{C} & =  \hat\sigma_+ \hat F + \hat\sigma_- \hat F^\dagger \, ,
\end{align}
where the fermion coupling operator is
\begin{align}\label{eq:FermionCouplingOperator}
\hat F &= \sum_{klm} T_{klm} \hat c^\dagger_{km} \hat d_{lm} \, .
\end{align}
Here the operators $\hat c^{(\dagger)}$ and $\hat d^{(\dagger)}$ annihilate (create) an electron in the system and in the bath, correspondingly.
Simultaneously, the state of the island spin is flipped.
The transverse level~$m$ does not change in this electron transport event.

It has been understood that
in the limit of a large number of transverse levels, 
the bath coupling operator~$\hat F$ becomes a Gaussian variable.
Physically, in this limit, consecutive spin-flips which are induced by
the system-bath electron hopping
always involve electron transitions corresponding to separate transverse levels.
On the Keldysh contour, this step is often referred to as the loop approximation, which is taken for example in Ref.~\cite{SchoellerSchon1994} for metallic transistors.
For systems with only one transverse channel instead, the loop approximation can be valid (at least)
when the bath memory time is shorter than the average time between electron hoppings between the island and a lead.

When establishing the equivalent spin-boson model, a central role is played by the spectral function of the coupling operator~$\hat F$,
\begin{align}\label{eq:SETspectrum}
&S(\omega) = \left\langle \hat F^\dagger(t) \hat F(0)\right\rangle_\omega = \left\langle \hat F(t) \hat F^\dagger(0) \right\rangle_\omega \nonumber \\
&= 2\pi\sum_m\sum_{k}\sum_{l} \vert T_{klm}\vert^2 \delta(\omega-E_{lm}/\hbar+E_{km}/\hbar  ) \nonumber\\
&\times f(E_{lm}) \left[1-f(E_{km})\right] \, ,
\end{align}
where $f(E)$ is the Fermi function
and the index~$\omega$ refers to a Fourier transformation under a free evolution of the bath, as in Eq.~(\ref{eq:multidimensional_spectral_function}).
When writing down the corresponding spin-boson model, we also need to account for the fact that the coupling operator $\hat F$ is not Hermitian and
that the opposite-direction spin-flips are described by the same spectral function~$S(\omega)$, see Eq.~(\ref{eq:SETspectrum}).
It follows that replacing $\hat F$ by a boson position-operator $\hat B + \hat B^\dagger$, or by $\hat B$, is not satisfactory.
A  replacement of type
\begin{align}
\hat F \rightarrow \hat B_1 +\hat B_2^\dagger \, .
\end{align}
where operators $\hat B^{(\dagger)}_1$ and $\hat B^{(\dagger)}_2$ are independent but characterized by a spectral function of the same form (see below),
is found to be satisfactory.
Since the free-evolution statistics of bosonic creation and annihilation operators are Gaussian,
the two baths (fermionic and bosonic) are indistinguishable for an observer from the system, if the spectral functions match.

According to these observations,
a spin-boson model that describes the original spin-fermion problem is of the form
\begin{align}
\hat H_\textrm{S} & = -\frac{\hbar\Delta}{2}\hat\sigma_z \, ,\\
\hat H_\textrm{B} &= \sum_{k} \hbar\omega_{k} \hat B^\dagger_{1k} \hat B_{1k} + \sum_{k} \hbar\omega_{k} \hat B^\dagger_{2k} \hat B_{2k} \, , \\
\hat H_\textrm{C} &=  \hat \sigma_+\sum_{k}v_{k}\left(\hat B_{1k} + \hat B_{2k}^\dagger\right) + \hat \sigma_-\sum_{k}v_{k}\left(\hat B_{1k}^\dagger + \hat B_{2k}\right)  \, . \label{eq:equivalent_Hamiltonian1}
\end{align}
A spin flip $\hat\sigma_+$ absorbs a boson from field~$1$ and creates a boson into field~$2$.
The spectral function, in thermal equilibrium, is given by:
\begin{align}\label{eq:multi_TLS2}
S(\omega) &= \sum_k \left\langle \hat B^\dagger_{1k}(t)\hat B_{1k}(0) \right\rangle_\omega + \sum_k \left\langle \hat B_{2k}(t)\hat B^\dagger_{2k}(0) \right\rangle_\omega \nonumber\\
&=2\pi \sum_k \frac{ v_{k}^2\delta(\vert\omega\vert - \omega_{k})} {1-\exp\left( -\frac{\hbar\omega}{k_\textrm{B}T} \right)}\textrm{sign}(\omega) \, .
\end{align}
The spin-boson model parameters $v_k$ are chosen so that the spectral function of the fermionic problem is reproduced.
In the example of Sec.~\ref{sec:SET},
we choose a structured spectral function of the form 
\begin{align}\label{eq:SETspectrum_example}
S(\omega) &= \sum_{k=0}^1\frac{\alpha\omega}{1 - \exp\left(-\frac{\omega} {k_\textrm{B} T}\right)} \frac{1}{2\pi} \frac{\kappa'}{\left(\frac{\kappa'}{2}\right)^2 + (\omega-\omega_k)^2}   \, ,
\end{align}
where $\alpha=0.25$, $\kappa'=0.4\omega_0$, $\omega_1=2\omega_0$, and
a cut-off function $1 / [1 + (\omega/\omega_\textrm{c})^4]$ with $\omega_\textrm{c}=\sqrt{3}\omega_0$.
The spectral function~$S(\omega)$ is then the target function in the coarse graining.

The form of coupling Hamiltonian~(\ref{eq:equivalent_Hamiltonian1}) supports the use of the variable iSWAP two-qubit decomposition,
since it is the native gate of (system-bath) coupling $\hat\sigma_+\hat\sigma_-^i + \hat\sigma_-\hat\sigma_+^i$.
The counter-rotating coupling terms $\hat\sigma_+\hat\sigma_+^i + \hat\sigma_-\hat\sigma_-^i$ can be created by surrounding the variable iSWAP by X-gates, see Appendix~\ref{Appendix:optimized_algorithms}.

Alternatively, an equivalent spin-boson Hamiltonian supporting the variable MS (native XX) decomposition
can be derived by defining new bosonic operators
\begin{align}\label{eq:hermitian coupling1}
\hat a_{k} &= \frac{1}{\sqrt{2}}\left(\hat B_{1k} + \hat B_{2k} \right) \\
\hat b_{k} &= \frac{\textrm{i}}{\sqrt{2}}\left(-\hat B_{1k} + \hat B_{2k} \right) \, .
\end{align}
This changes the form of the coupling Hamiltonian,
\begin{align}
H_\textrm{B} &= \sum_{k} \hbar\omega_{k} \hat a^\dagger_{k} \hat a_{k} + \sum_{k} \hbar\omega_{k} \hat b^\dagger_{k} \hat b_{k} \, , \\
H_{\rm C} &= \hat \sigma_x\sum_k \frac{v_k}{\sqrt{2}}\left(\hat a_k^\dagger + \hat a_k\right) + \hat \sigma_y\sum_k \frac{v_k}{\sqrt{2}}\left(\hat b_k^\dagger + \hat b_k\right)  \, . \label{eq:equivalent_Hamiltonian2}
\end{align}
When representing this with spin-spin Hamiltonian, the first coupling-term is replaced by $\hat\sigma_x\hat\sigma_x^i$-interaction
and the second coupling-term by $\hat\sigma_y\hat\sigma_x^i$-interaction.
The first coupling type is native to variable MS, whereas the second one can be created by surrounding the variable MS by $\pi/2$ Z-rotations, see Appendix~\ref{Appendix:optimized_algorithms}.
It follows that the circuit depth, when time-propagating the system with variable MS according to this Hamiltonian,
will be the same as when time-propagating the system with variable iSWAP according to the Hamiltonian~(\ref{eq:equivalent_Hamiltonian1}).

When time-evolving the open-system model on the quantum computer with a noisy system qubit,
system-qubit X-gates are introduced between Trotter steps. This is done to transform physical system-qubit decay partly into effective excitation,
see Appendix~\ref{Appendix:optimized_algorithms}.
After this symmetrization, the physical decay noise corresponds to an effective noise that has equal contribution of system decay and excitation.
Equivalently, the physical decay noise corresponds to an effective noise that has equal contribution of incoherent $X$-flips and $Y$-flips,
see Eq.~(\ref{eq:symmetrization}).
These system-environment coupling operators are the same as in the considered system-bath model.
It follows that system noise can be mapped to a constant background of the bath the spectral functions, see Sec.~\ref{sec:coarse_graining_scheme2} and Eq.~(\ref{eq:spectral_function_with_background}).
This mapping is demonstrated in practice in the example of Sec.~\ref{sec:SET}.

\bibliography{spinboson}

\end{document}